\documentclass[aps,prl, twocolumn,superscriptaddress,nobibnotes]{revtex4-2}

\usepackage{amsmath,amsthm,amsfonts,amssymb,amscd} 
\usepackage{graphicx}
\usepackage{xcolor}
\usepackage[T1]{fontenc}
\usepackage{here}
\usepackage{enumerate}
\usepackage{braket}
\usepackage{url}
\usepackage[normalem]{ulem}
\usepackage{soul}
\usepackage{cancel}
\usepackage{mathtools}
\usepackage{commath}
\usepackage[colorlinks=true,citecolor=teal,linkcolor=red!70!white,urlcolor=blue!70!white,hyperindex,breaklinks]{hyperref}
\usepackage{thmtools,thm-restate}
\usepackage{placeins}

\newcommand{\add}[1]{{\color{blue}#1}}

\newcommand{\todai}{Department of Physics, Graduate School of Science, The University of Tokyo, Hongo 7-3-1, Bunkyo-ku, Tokyo 113-0033, Japan}

\newcommand{\transscale} {Trans-scale Quantum Science Institute, The University of Tokyo, Hongo 7-3-1, Bunkyo-ku, Tokyo 113-0033, Japan}
\newcommand{\perimeter}{Perimeter Institute for Theoretical Physics, 31 Caroline Street North, Waterloo, Ontario, N2L 2Y5, Canada}
\newcommand{\iqc}{Institute for Quantum Computing, University of Waterloo, 200 University Avenue West, Waterloo, Ontario, N2L 3G1, Canada}
\newcommand{\geneve}{Department of Applied Physics, University of Geneva, 1205 Geneva, Switzerland}
\newcommand{\sorbonne}{Sorbonne Universit\'{e}, CNRS, LIP6, F-75005 Paris, France}

\newtheorem{theorem}{Theorem}%[chapter]
\newtheorem*{theorem*}{Theorem}
\newtheorem{lemma}[theorem]{Lemma}%[chapter]
\newtheorem{conjecture}{Conjecture}% [chapter]
\newcommand{\dket}[1]{\vert#1\rangle\!\rangle}
\newcommand{\dbra}[1]{\langle\!\langle #1\vert}

\newcommand{\dketbra}[2]{\left.\left\vert #1 \right\rangle \hspace{-.8mm} \right\rangle \hspace{-1mm} \left\langle\hspace{-.8mm}\left\langle #2 \right\vert\right.}

\def\d{{\rm d}}

\newcommand{\map}[1]{\mathcal{#1}}
\newcommand{\supermap}[1]{\map{#1}}
\newcommand{\switch}{\mathtt{SWITCH}}
\newcommand{\Tr}{\operatorname{Tr}}

\usepackage{dsfont}
\usepackage{physics}

\newcommand{\1}{\mathds{1}}

\newcommand{\mcA}{\mathcal{A}}
\newcommand{\mcB}{\mathcal{B}}
\newcommand{\mcC}{\mathcal{C}}

\newcommand{\mcH}{\mathcal{H}}

\newcommand{\mcS}{\mathcal{S}}

\newcommand{\mbA}{\mathbb{A}}
\newcommand{\mbB}{\mathbb{B}}
\newcommand{\mbC}{\mathbb{C}}

\newcommand{\mbL}{\mathbb{L}}

\makeatletter
\def\dbraket#1{%
    \@ifnextchar\bgroup{%
        \dbraket@{#1}%
    }{%
        \langle\!\langle {#1} \vert {#1} \rangle\!\rangle%
    }%
}
\def\dbraket@#1#2{%
    \langle\!\langle {#1} \vert {#2} \rangle\!\rangle%
}
\makeatother

\makeatletter
\def\dketbra#1{%
    \@ifnextchar\bgroup{%
        \dketbra@{#1}%
    }{%
        \vert {#1} \rangle\!\rangle\!\langle\!\langle {#1} \vert%
    }%
}
\def\dketbra@#1#2{%
    \vert {#1} \rangle\!\rangle\!\langle\!\langle {#2} \vert%
}
\makeatother

\usepackage{listings}
\begin{document}
 
\title{Exponential separation in quantum query complexity of the quantum switch with respect to simulations with standard quantum circuits}

\author{Hl\'er Kristj\'ansson}
\thanks{These three authors contributed equally. \\ Correspondence to: \href{mailto:hler.kristjansson@outlook.com}{hler.kristjansson@outlook.com} \\ and \href{mailto:satoshiyoshida.phys@gmail.com}{satoshiyoshida.phys@gmail.com}}
\affiliation{\perimeter}
\affiliation{\iqc}
\affiliation{\todai}
\author{Tatsuki Odake}
\thanks{These three authors contributed equally. \\ Correspondence to: \href{mailto:hler.kristjansson@outlook.com}{hler.kristjansson@outlook.com} \\ and \href{mailto:satoshiyoshida.phys@gmail.com}{satoshiyoshida.phys@gmail.com}}
\affiliation{\todai}
\author{Satoshi Yoshida}
\thanks{These three authors contributed equally. \\ Correspondence to: \href{mailto:hler.kristjansson@outlook.com}{hler.kristjansson@outlook.com} \\ and \href{mailto:satoshiyoshida.phys@gmail.com}{satoshiyoshida.phys@gmail.com}}
\affiliation{\todai}
\author{\mbox{Philip Taranto}}
\affiliation{\todai}
\author{Jessica Bavaresco}
\affiliation{\geneve}
\author{Marco T\'{u}lio Quintino}
\affiliation{\sorbonne}
\author{Mio Murao}
\affiliation{\todai}
\affiliation{\transscale}

\begin{abstract}
Quantum theory is consistent with a computational model permitting black-box operations to be applied in an indefinite causal order, going beyond the standard circuit model of computation. The quantum switch---the simplest such example---has been shown to provide numerous information-processing advantages. Here, we prove that the action of the quantum switch on two $n$-qubit quantum channels \emph{cannot} be simulated deterministically and exactly by any causally ordered quantum circuit that uses $M$ calls to one channel and one call to the other, if $M \leq \max(2, 2^n-1)$. 
This demonstrates an exponential separation in quantum query complexity of indefinite causal order compared to standard quantum circuits.
\end{abstract}

\maketitle

\textit{Introduction.---}The possibility of performing quantum operations in an indefinite causal order has attracted significant attention~\cite{hardy2005probability,hardy2009quantum,chiribella2009beyond,chiribella2013quantum,oreshkov2012quantum,baumeler2014maximal,baumeler2016space,araujo2017purification,goswami2018indefinite,wechs2021quantum}. From a foundational point of view, this possibility 
has profound consequences for understanding  causality and deep implications for the quantum nature of space-time \cite{hardy2005probability,hardy2009quantum, zych2019bell,paunkovic2020causal}. From an information-processing 
perspective, it is equally significant, challenging the standard conception of computation in which operations are performed in a fixed order on a system \cite{chiribella2009beyond,chiribella2013quantum,colnaghi2012quantum,baumeler2014maximal,baumeler2016space}. The simplest example of a process with indefinite causal order is the \emph{quantum switch}, a transformation that takes a single call to each of two quantum channels $\map A$ and $\map B$ as input, and returns a superposition \cite{chiribella2019quantum} of their two possible orderings $\map B \circ \map A$ and $\map A \circ \map B$, conditioned on the state of a control qubit \cite{chiribella2009beyond,chiribella2013quantum}.

The ability to perform operations in such an indefinite order has been shown to provide advantages in a variety of information-processing settings, including quantum query complexity \cite{araujo2014computational,araujo2017quantum,renner2021reassessing,renner2022computational,abbott2023quantum,taddei2021computational}, quantum communication complexity \cite{guerin2016exponential}, multipartite games \cite{oreshkov2012quantum,baumeler2014maximal,baumeler2016space}, quantum Shannon theory \cite{ebler2018enhanced,salek2018quantum,chiribella2018indefinite,loizeau2020channel,procopio2019communication,procopio2020sending,sazim2021classical,chiribella2021quantum,goswami2020increasing,rubino2017experimental,rubino2021experimental,guo2020experimental}, quantum metrology \cite{zhao2020quantum,chapeau2021noisy,liu2023optimal,yin2023experimental}, quantum channel discrimination~\cite{chiribella2012perfect,bavaresco2021strict,bavaresco2022unitary}, and quantum thermodynamics \cite{felce2020quantum,nie2022experimental}, most of which are due to the quantum switch. While the realization of indefinite causal order within the framework of known physics \cite{procopio2015experimental,goswami2018indefinite,goswami2020increasing,rubino2017experimental,rubino2021experimental,guo2020experimental} or in potential future theories of quantum gravity \cite{zych2019bell,paunkovic2020causal} remains a matter of debate \cite{araujo2017purification,chiribella2019quantum,kristjansson2021witnessing,oreshkov2019time,paunkovic2020causal,vilasini2022embedding,ormrod2023causal}, these information-theoretic advantages have garnered independent interest,  motivated by fundamental concerns in information theory and computation.

In the context of quantum computation, whether the quantum switch exhibits a true complexity-theoretic advantage depends upon whether its action can be efficiently simulated by using causally ordered quantum circuits, given extra queries to one (or both) of the channels. Until now, no exponential separation has been demonstrated between the query complexity of computations using indefinite causal order versus standard quantum circuits. Indeed, for unitary channels, the quantum switch can be simulated by a quantum circuit with a fixed order and just one extra query \cite{chiribella2013quantum}, significantly limiting the computational power of the quantum switch in the case of unitary inputs. A crucial open question is whether this limitation extends to general quantum channels. 

In this Letter, we answer this in the negative by proving a no-go theorem: the quantum switch of two $n$-qubit channels cannot be deterministically and exactly simulated by a quantum circuit with fixed causal order (or classically controlled causal order; see below), with one call to one channel and $M$ calls to the other, as long as $M\leq \max(2, 2^n-1)$. We further conjecture that a similar bound holds for $M$ calls to both channels. Our theorem demonstrates an exponential separation in quantum query complexity for computational tasks using quantum processes with indefinite causal order versus standard quantum circuits (as well as those with classical control of causal order), in terms of the number of qubits. If our conjecture holds, it would imply that processes with indefinite causal order cannot be efficiently simulated using standard quantum circuits (or even with classically controlled causal order).

\textit{Framework.---}Quantum processes with indefinite causal order arise as a special case of higher-order quantum transformations \cite{chiribella2008quantum,chiribella2013quantum,chiribella2018indefinite} (also known as quantum supermaps \cite{chiribella2008transforming} or process matrices \cite{oreshkov2012quantum}). Higher-order quantum transformations are defined according to the following hierarchy. We denote the set of linear operators on a finite-dimensional Hilbert space $\mathcal{H}^A$ corresponding to a physical system $A$ as $\mbL(A)$. 
A \textit{quantum state} is any linear operator $\rho \in \mbL(A)$ that is positive semidefinite $\rho\ge 0$ and of unit trace $\Tr [\rho] = 1$. A \textit{quantum channel} is any consistent map from quantum states to quantum states, i.e.,\ any linear map $\map C: \mbL(I) \to \mbL(O)$ that is both \textit{completely positive} (\textbf{CP}) and \textit{trace preserving} (\textbf{TP}). A \textit{quantum supermap} $\map S$ is any consistent map from the space of $M$-tuples of quantum channels to the space of quantum channels. Mathematically, the consistency condition for supermaps  requires that an $M$-slot quantum supermap is any $M$-linear map $\map S: \bigotimes_{i=1}^{M}[\mbL(I_i) \to \mbL(O_i)] \to [\mbL(P) \to \mbL(F)]$ that is both completely CP-preserving and TP-preserving \cite{quintino2019probabilistic,gour2019comparison}. 

Throughout, we will use the Choi representation \cite{choi1975completely,jamiolkowski1972linear} of quantum transformations. Any linear operator $V: \mathcal{H}^A \to \mathcal{H}^B$ can be represented by its \textit{Choi vector}
\begin{equation}
	\dket{V} := \sum_{i} \ket{i}^A \otimes V \ket{i}^A \in  \mathcal{H}^A \otimes \mathcal{H}^B \, .
\end{equation}
Similarly, any linear map $\map Q: \mbL(A) \to \mbL(B)$ is isomorphic to its \textit{Choi matrix} 
\begin{equation}
	Q :=  \sum_{ij} \ketbra{i}{j}^A \otimes \map Q (\ketbra{i}{j}^A ) \in \mbL(A \otimes B) \, .
\end{equation}
In both cases, $\{\ket{i}\}_i$ are computational basis vectors. For clarity, we use calligraphic letters $\map Q$ for linear maps and standard font $Q$ for the associated Choi matrix.

Any quantum channel $\map C: \mbL(I) \to \mbL(O)$ corresponds to a positive semidefinite Choi matrix $C \in \mbL(I \otimes O)$ normalized such that $\Tr_{O} [C] = \1_I$, where $\1_I$ is the identity matrix on $\mathcal{H}^I$. Similarly, any quantum supermap $\map S: \bigotimes_{i=1}^{M}[\mbL(I_i) \to \mbL(O_i)] \to [\mbL(P) \to \mbL(F)]$ has a Choi representation as a positive semidefinite matrix $S \in \mbL[P \otimes (\bigotimes_{i=1}^{M}I_i \otimes O_i) \otimes F]$, called the \textit{process matrix}, that is restricted to a specific subspace (corresponding to TP-preservation) and normalized such that $\Tr [S] = d^P \Pi_{i=1}^M d^{I_i}$, where $d^A := \mathrm{dim}(\mcH^A)$ \cite{araujo2015witnessing}.

The composition of quantum states, channels, and supermaps is calculated in the Choi representation via the \emph{link product} $\ast$ \cite{chiribella2009theoretical}. For any two matrices $Q \in \mbL(A\otimes B), R \in \mbL(B \otimes C)$, the link product is defined as $Q * R := \Tr_B [( Q^{AB} \otimes \1^C)^{{\rm T}_B} (\1^A \otimes R^{BC} ) ]$, with $\rm T_B$ representing the partial transpose with respect to system $B$. In particular, the action of a quantum supermap $\mcS$ on a set of quantum channels $\{ \map C_1, \hdots, \map C_M\}$ is given by $\map S (\map C_1 , \ldots , \map C_M) := \map S (\map C_1 \otimes \ldots \otimes \map C_M)$, which, in Choi operator form, is equivalent to $S * (C_1 \otimes \cdots \otimes C_M)$.

\begin{figure}
    \centering
    \includegraphics[width=\linewidth]{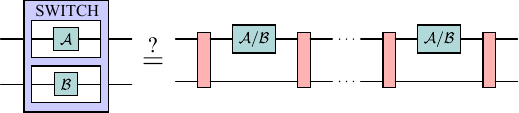}
    \caption{We consider the question of simulating the action of the quantum switch on two black-box quantum channels $\mcA$ and $\mcB$ (left) using a quantum circuit with fixed causal order (QC-FO) (right) or a quantum circuit with classical control of causal order (QC-CC) \cite{supple}.
    In the QC-FO shown on the right, $\mcA$ or $\mcB$ are called sequentially $M$ and $N$ times, respectively. }
    \label{fig:quantum_comb_simulation}
\end{figure}

Ordinary quantum circuits correspond to the special class of quantum supermaps known as quantum combs \cite{chiribella2008quantum} or \textit{quantum circuits with fixed causal order} (\textbf{QC-FOs}) \cite{wechs2021quantum}, which can be realized by a sequence of quantum gates, interspersed with open slots.
An $M$-slot quantum circuit with fixed causal order is a quantum supermap $\map S$ that can be decomposed as a quantum circuit with $M+1$ fixed quantum channels $\map V_0 :  \mbL(P) \to \mbL(I_1 \otimes E_1), \map  V_1 : \mbL(O_1 \otimes E_1) \to \mbL(I_2 \otimes E_2), \dots, \map V_M :  \mbL(O_M \otimes E_M) \to \mbL(F)$, connected sequentially with auxiliary systems $\{E_i\}_{i=1}^M$. In the Choi representation, this is equivalent to $S = V_M \ast \cdots \ast V_0$. The action of such a supermap on $M$ input quantum channels $\{ C_{i} : \mbL(I_i) \to \mbL(O_i) \}_{i=1}^M$ inserted into the slots between each $\map V_i$ is given by $S * (C_1 \otimes \cdots \otimes C_M) = V_M * C_M * \cdots * V_1 * C_1 * V_0 $. 

However, QC-FOs are not the most general quantum supermaps that can be considered to have an underlying definite causal structure. Convex combinations of QC-FOs and quantum supermaps where the order of operations is determined dynamically are also possible. A more general class of transformations that includes such possibilities is \textit{quantum circuits with classical control of causal order} (\textbf{QC-CCs}) \cite{wechs2021quantum}, whose characterization is given in the Supplemental Material (\textbf{SM}), Lemma \ref{lem:qccc_invalid} \cite{supple}. QC-CCs encompass the most general transformations known to be achievable by standard quantum computers operating in a definite causal order. As such, any computational advantage of processes with indefinite causal order is most reasonably determined by comparison with QC-CCs (which include the standard QC-FOs)
\footnote{Formally, quantum supermaps that are not compatible with an underlying definite causal structure are called \textit{causally non-separable processes} \cite{oreshkov2012quantum,araujo2015witnessing,wechs2019definition} and are said to have indefinite causal order. Although it is currently an open question whether there exist causally separable processes that are not QC-CCs \cite{wechs2021quantum}, no such processes have been found to date, and therefore, any computational advantage of processes with indefinite causal order remains most reasonably determined by comparison with QC-CCs.}. 

\textit{Query complexity of higher-order quantum transformations.---}We study the following type of tasks. Consider a classical description of a function $f: [\mbL(I) \to \mbL(O)] \otimes [\mbL(I') \to \mbL(O')] \to [\mbL(P) \to \mbL(F)]$ which takes a pair of quantum channels $\mcA, \mcB$ as inputs to an output quantum channel $f(\mcA,\mcB)$. We say that a quantum supermap $\map S$ \textit{simulates} the function $f$ deterministically and exactly \footnote{In contrast, a probabilistic simulation would have an equality with probability $p$, and an approximate simulation would have an approximate equality with error $\epsilon$.} if, given $M$ black-box queries to the quantum channel $\map A$ and $N$ black-box queries to the quantum channel $\map B$, $\map S(\map A^{\otimes M}, \map B ^{\otimes N} ) = f(\map A, \map B)$. See Fig.~\ref{fig:quantum_comb_simulation} for a graphical depiction of simulating the action of the quantum switch using a QC-FO supermap.

In general, the number of calls to each of the input channels is a fundamental resource to the simulability of a function. In the case where one of the channels is fixed to being called $N=1$ times, we can define a simple notion of quantum query complexity that depends only on the number of calls to the other channel, $M$. We define the \textit{one-sided quantum query complexity} of a function $f$, with respect to a class of supermaps $\mathbb{S}$, as the minimum number of queries $M$ while $N=1$, over all supermaps $\map S \in \mathbb{S}$ such that $\map S$ simulates $f$. This definition can be seen as a step towards a fully quantum generalization of the notion of query complexity. While the standard notion of quantum query complexity has so far typically been defined for classical (e.g., boolean) functions, here we consider the query complexity of functions whose inputs and outputs are themselves quantum channels (see also \cite{odake2024analytical}). This is similar in spirit to recent works on the complexity of preparing quantum states \cite{rosenthal2021interactive,metger2023stateqip}.

\textit{Simulating the quantum switch}.---The simplest and most widely studied example of a process with indefinite causal order is the \textit{quantum switch} \cite{chiribella2013quantum}. The quantum switch combines two quantum channels $\map A : [\mbL(I) \to \mbL(O)]  $ and $\map B : [\mbL(I') \to \mbL(O')]$ in two possible sequential orderings, depending on the quantum state of a control qubit $P_C$.
The process matrix of the $n$-qubit quantum switch $\map S_\switch :  [[\mbL(I) \to \mbL(O)] \otimes  [\mbL(I') \to \mbL(O')]]  \to [\mbL(P_C \otimes P_T) \to \mbL(F_C \otimes F_T)]$, where $I, O, I', O', P_T, F_T$ correspond to $n$-qubit Hilbert spaces and $P_C, F_C$ correspond to qubit Hilbert spaces, is given by
 $S_\switch = \dketbra{S_\switch}$, with 
\begin{align}
	\dket{S_\switch}^{PFIO I' O'} 
	&\coloneqq \ket{0}^{P_C}\ket{0}^{F_C}\dket{\1}^{P_T I}\dket{\1}^{O I'}\dket{\1}^{O'F_T} \nonumber \\
	&\hspace{-2em}+ \ket{1}^{P_C}\ket{1}^{F_C}\dket{\1}^{P_T I'}\dket{\1}^{O'I}\dket{\1}^{O F_T}  .
\end{align}
In the case where the input channels are unitary, i.e.,\ $\map U (\cdot) = U (\cdot) U^\dag$ and $\map V (\cdot) = V (\cdot) V^\dag$ for some unitary operators $U,V$, the action of the quantum switch takes the simple form $\map S_\switch (\map U, \map V) (\cdot) = S_\switch (\cdot)S_\switch^\dag $, with
\begin{equation}\label{eq:switch_on_U}
	\begin{split}
 S_\switch  =
 VU \otimes \ketbra{0}{0} +  UV \otimes \ketbra{1}{1}   .
\end{split}
\end{equation}

\begin{figure}
    \centering
    \includegraphics[width=0.8\linewidth]{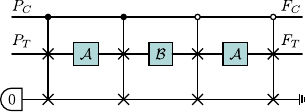}
    \caption{A quantum circuit with fixed causal order taking two calls to a quantum channel $\map A$ and one call to a quantum channel $\map B$. This circuit simulates the action of the quantum switch on any unitary channel $\map A$ and quantum channel $\map B$.}
    \label{fig:switch_simulation}
\end{figure}

To understand the computational power of the quantum switch, it is essential to know whether its action can be efficiently simulated with causally ordered quantum supermaps by using more queries to one or both of the input channels. The one-sided quantum query complexity of the function $\map S_\switch$ with respect to the set of all (including indefinite causal order) supermaps is trivially 1. The action of the quantum switch on two unitary channels $\map U, \map V$ can be simulated deterministically and exactly with a quantum circuit of fixed causal order $\map C_{\rm sim}$ using just one extra call to either of the two channels \cite{chiribella2013quantum}:
\begin{equation}
	\map C_{\rm sim}( \map U, \map V, \map U) = \map S_\switch(\map U, \map V) \quad \forall \,  \map U, \map V \, .
\end{equation}
This result holds for any size of the target system. The circuit for $\map C_{\rm sim}$ is depicted in Fig. \ref{fig:switch_simulation}. 

Interestingly, we observe that the same quantum circuit $\mcC_{\rm sim}$ can simulate the action of the quantum switch on one unitary channel and one general quantum channel, with only one extra call to the unitary channel. That is, for any unitary channel $\map U$ and any quantum channel $\map B$, we have
\begin{equation}
	\map C_{\rm sim}( \map U, \map B, \map U) = \map S_\switch(\map U, \map B) \quad \forall \,  \map U, \map B \, .
\end{equation}
However, whenever $\mcC_{\rm sim}$ is applied to a pair of general quantum channels $\map A, \map B$ (with, e.g., two copies of $\map A$) it does \textit{not} reproduce the action of the quantum switch. 

\textit{No-go theorem.---}Naturally, one might wonder whether there exists some other causally ordered supermap---either a QC-FO or QC-CC---that can reproduce the action of the quantum switch on general quantum channels given $M\ge 2$ queries to one of the two $n$-qubit channels. Here, we answer this in the negative for $M \leq \max(2, 2^n-1)$.
\begin{restatable}{theorem}{main}\label{theo:main}
There is no $(M+1)$-slot supermap $\map{C}$, for $M \leq  \max(2, 2^{n}-1)$, with fixed causal order or classical control of the causal order, satisfying
\begin{align}
	\supermap{C}( \underbrace{\map{A},\dots, \map{A}}_M, \map{B} ) &= \map S_\switch(\map{A}, \map{B})\label{eq:switch_simulation_1_maintext}
\end{align}
for all $n$-qubit mixed unitary channels $\map{A}$ and unitary channels $\map{B}$.
 
Therefore, such a supermap also does not exist for all $n$-qubit quantum channels $\map{A}$ and $\map{B}$.
\end{restatable}

\noindent This implies that the one-sided quantum query complexity of the action of the quantum switch, with respect to all causally ordered supermaps, is lower-bounded by $\max(3,2^n)$.

\begin{proof}
We provide the full proof in the SM~\cite{supple}; here, we give a sketch of the proof for the case where $M=2$, which is shown by contradiction. Let $\map C : \bigotimes_{i=1}^{2}[\mbL(I_i) \to \mbL(O_i)] \otimes [\mbL(I'_1) \to \mbL(O'_1)] \to [\mbL(P_C \otimes P_T) \to \mbL(F_C \otimes F_T)]$ be the 3-slot QC-CC quantum supermap that simulates the action of the quantum switch on all mixed unitary quantum channels. 
For arbitrary unitary channels $\map{U}_1, \map{U}_2, \map{V}$, the supermap $\map{C}$ necessarily respects
\begin{align}
\hspace{-0.2em}\forall \, l\in\{1,2\}: \quad 
    \map{C}(\map{U}_l, \map{U}_l, \map{V}) &= \map{S}_\switch(\map{U}_l, \map{V})  , \label{eq:C_linear_unitary_2}\\
    \hspace{-0.2em}\map{C}\left({\map{U}_1+\map{U}_2 \over 2}, {\map{U}_1+\map{U}_2 \over 2}, \map{V} \right) &= \map{S}_\switch \left({\map{U}_1+\map{U}_2 \over 2}, \map{V} \right) . \label{eq:C_linear_unitary}
\end{align}
By linearity, Eqs.~\eqref{eq:C_linear_unitary_2} and \eqref{eq:C_linear_unitary} imply that
\begin{align}
    \map{C}(\map{U}_1, \map{U}_2, \map{V}) + \map{C}(\map{U}_2, \map{U}_1, \map{V}) = \map{S}_\switch(\map{U}_1+\map{U}_2, \map{V}) \,.
\end{align}
Since $\map{C}(\map{U}_2, \map{U}_1, \map{V})$ is a CP map, $\map{S}_\switch(\map{U}_1 + \map{U}_2, \map{V}) - \map{C}(\map{U}_1, \map{U}_2, \map{V})$ is also CP. In terms of Choi matrices, this implies that
\begin{align}
    &C \ast (\dketbra{U_1}\otimes \dketbra{U_2}\otimes \dketbra{V})\\
    &\leq \dketbra{S_\switch}\ast [(\dketbra{U_1}+\dketbra{U_2})\otimes \dketbra{V}]. \nonumber
\end{align}

\noindent Since $\map{C}$ is a QC-CC by assumption, its Choi matrix $C$ can be decomposed as $C = \sum_{(i,j,k)\in\mathrm{Perm}(1,2,3)}C_{ijk}$
such that $C_{ijk}$ satisfies $C_{ijk}\geq 0$ and several affine conditions \cite{wechs2021quantum} (which we call the \textit{QC-CC conditions}; see  Lemma \ref{lem:qccc_invalid} in the SM~\cite{supple}). Using an eigendecomposition of $C_{ijk}$ given by $C_{ijk} = \sum_a \dketbra{C^{(a)}_{ijk}}$,
it follows that
\begin{align}
\label{eq:inequality_Cijka}
    &\dketbra{C^{(a)}_{ijk}}\ast(\dketbra{U_1}\otimes \dketbra{U_2}\otimes \dketbra{V})\\
    &\leq \dketbra{S_\switch}\ast [(\dketbra{U_1}+\dketbra{U_2})\otimes \dketbra{V}],\nonumber
\end{align}
for all $i,j,k$ and $a$. 

Defining the link product $\ast$ for Choi vectors $\dket{Q}\in \mcH^A\otimes \mcH^B$ and $\dket{R}\in \mcH^B\otimes \mcH^C$ as $\dket{Q} \ast \dket{R} \coloneqq \sum_i (\1^A \otimes \bra{i}^{B})\dket{Q} \otimes (\bra{i}^B \otimes \1^C)\dket{R}$ using the computational basis $\{\ket{i}\}_i$ on $\mcH^B$, $\dketbra{Q} \ast \dketbra{R}$ is given by $(\dket{Q}\ast \dket{R}) (\dket{Q}\ast \dket{R})^\dagger$ (see, e.g., Lemma 1 of Ref.~\cite{yokojima2021consequences}).
Thus, the support of the right-hand side of Eq.~\eqref{eq:inequality_Cijka} is given by $\mathrm{span}\{\dket{S_\switch} \ast (\dket{U_l} \otimes \dket{V})\}_{l=1}^{2}$ and that of the left-hand side of Eq.~\eqref{eq:inequality_Cijka} is given by the projector onto a one-dimensional subspace spanned by $\dket{C^{(a)}_{ijk}}\ast (\dket{U_1}\otimes \dket{U_2}\otimes \dket{V})$.
Therefore, one can write
\begin{align}
\label{eq:span}
    &\dket{C^{(a)}_{ijk}}\ast (\dket{U_1}\otimes \dket{U_2}\otimes \dket{V})  \nonumber\\
    &= \sum_{l=1}^{2} \xi_{ijk}^{(a, l)} (U_1, U_2, V) \dket{S_\switch}\ast (\dket{U_l}\otimes \dket{V}),
\end{align}
for some $\xi_{ijk}^{(a, l)} (U_1, U_2, V) \in \mathbb{C}$. A proof of this fact is in Lemma~\ref{lem:span} in the SM. In Lemma~\ref{lem:linearity_MN} in the SM, we generalize this result for $M>2$.

We now invoke Lemma~\ref{lem:indep_M1} and Lemma~\ref{lem:p_as_vector_MN}  in the SM to ensure that, when Eq.~\eqref{eq:span} is satisfied, there exist vectors $\dket{\xi_{ijk}^{(a, 1)}}\in \mcH^{I_2}\otimes \mcH^{O_2}$ and $\dket{\xi_{ijk}^{(a, 2)}}\in \mcH^{I_1}\otimes \mcH^{O_1}$, such that
\begin{align}
\label{eq:Ca}
    \dket{C^{(a)}_{ijk}} &= \sum_{l=1}^{2}\dket{S_\switch}^{PI_l O_l I_3 O_3 F} \otimes \dket{\xi_{ijk}^{(a, l)}},
\end{align}
for all $i,j,k$ and $a$, where $\dket{\xi_{ijk}^{(a, 1)}}$ and $\dket{\xi_{ijk}^{(a, 2)}}$ are independent of $U_1, U_2, $ and $V$.
Next, we argue why this is the case.

The basic idea for this part of the proof is based on differentiation with respect to a parametrization of the input unitary operators, a technique introduced concurrently in Ref.\ \cite{odake2024analytical} by some of the present authors. Suppose that $U_1,\ U_2,$ and $V$ are taken from the set $\{I,X,Y,Z\}$ of Pauli operators. 
If $U_1 \neq U_2$, then $\dket{ S_{\switch}}*(\dket{U_1}\otimes \dket{V})$ and $\dket{ S_{\switch}}*(\dket{U_2}\otimes \dket{V})$ are linearly independent. In this case, we can show using linearity that $\xi_{ijk}^{(a, 1)}(U_1, U_2, V)$ is independent of $U_1, V$ and $\xi_{ijk}^{(a, 2)}(U_1, U_2, V)$ is independent of $U_2, V$.  If on the other hand $U_1 = U_2 = \sigma$, then $\dket{ S_{\switch}}*(\dket{U_1}\otimes \dket{V})$ and $\dket{ S_{\switch}}*(\dket{U_2}\otimes \dket{V})$ are \textit{not} linearly independent.

In such cases, it turns out that $\xi_{ijk}^{(a, 1)}(\sigma, \sigma, V)$ and $\xi_{ijk}^{(a, 2)}(\sigma, \sigma, V)$ can be suitably chosen as $\xi_{ijk}^{(a, 1)}(\sigma', \sigma, V)$ and $\xi_{ijk}^{(a, 2)}(\sigma, \sigma', V)$, respectively, where $\sigma' \neq \sigma$ is a Pauli operator.
Note that $\xi_{ijk}^{(a, 1)}(\sigma', \sigma, V)$ and $\xi_{ijk}^{(a, 2)}(\sigma, \sigma', V)$ do not depend on the choice of $\sigma'$ as long as $\sigma'\neq \sigma$ holds.
The fact that such a redefinition is consistent with Eq.\ \eqref{eq:span} can be proven by differentiating the expression $\xi_{ijk}^{(a, l)}(\tilde{\sigma}(\theta), \tilde{\sigma}(\theta), V)$, where $\tilde{\sigma}(\theta)$ is a parameterized unitary operator satisfying $\tilde{\sigma}(0)=\sigma$ and $\frac{\rm d}{{\rm d}\theta}\!\mid_{\theta=0}\tilde{\sigma}(\theta)\propto\sigma'$. 
This redefinition implies that that $\xi_{ijk}^{(a, 1)}(U_1, U_2, V)$ and $\xi_{ijk}^{(a, 2)}(U_1, U_2, V)$ are independent of $U_1$ and $U_2$, respectively. By linearity, we can show that for $l \in 
\{1,2\}$, $\xi_{ijk}^{(a, l)}(U_1, U_2, V)$ is independent of $V$. 

The independence relations above imply that we can write $\xi_{ijk}^{(a, 1)} (U_1, U_2, V) = \dket{\xi_{ijk}^{(a, 1)}} \ast \dket{U_2}$ and  $\xi_{ijk}^{(a, 2)} (U_1, U_2, V) = \dket{\xi_{ijk}^{(a, 2)}} \ast \dket{U_1}$ for some vectors $\dket{\xi_{ijk}^{(a, 1)}}, \dket{\xi_{ijk}^{(a, 2)}}$. Substituting this into Eq.\ \eqref{eq:span} gives
\begin{align}
    &\dket{C^{(a)}_{ijk}}\ast (\dket{U_1}\otimes \dket{U_2}\otimes \dket{V}) \\
    &= \!\! \sum_{l=1}^{2}\dket{S_\switch}^{P I_l O_l I_3 O_3 F} \! \otimes \dket{\xi_{ijk}^{(a, l)}} \! \ast \! (\dket{U_1} \! \otimes \! \dket{U_2} \! \otimes \! \dket{V}) \, . \nonumber
\end{align}
Since this holds for all combinations of Pauli operators $U_1, U_2, V$, we obtain Eq.~\eqref{eq:Ca}. 

Hence, we have shown that a QC-CC simulation of the quantum switch implies the existence of vectors $\dket{\xi_{ijk}^{(a, 1)}}$ and $\dket{\xi_{ijk}^{(a, 2)}}$ such that Eq.~\eqref{eq:Ca} holds. Finally, we invoke Lemma~\ref{lem:qccc_invalid} in the SM, which states that supermaps with an eigendecomposition given by Eq.~\eqref{eq:Ca} \textit{cannot} satisfy the QC-CC conditions. This is a contraction, since we initially assumed that the supermap $\map{C}$ is QC-CC.
\end{proof}

\textit{Discussion.---}One might wonder whether, instead, there exists a supermap with $M \leq \max(2, 2^{n}-1)$ queries to $\map A$ and $N \leq \max(2, 2^{n}-1)$ queries to $\map B$ that could simulate the action of the quantum switch. Although the answer to this question is currently unknown, we conjecture that such a simulation is also impossible. 
\begin{conjecture} 
There is no $(M+N)$-slot supermap $\map{C}$ with fixed causal order or classical control of the causal order  satisfying
	\begin{align}
		\supermap{C}( \underbrace{\map{A},\dots, \map{A}}_M, \underbrace{\map{B},\dots,\map{B}}_N ) &= \map S_\switch(\map{A}, \map{B})\label{eq:switch_simulation_1_maintext}
	\end{align}
	for all  n-qubit quantum channels $\map{A}$ and $\map{B}$, if $\max(M,N) \leq  g(n)$, for some $g = \Theta(2^n)$ .
\end{conjecture}

\noindent Our rationale behind conjecturing that no simulation is possible even with multiple (albeit a sub-exponential number of) calls to \textit{both} channels is the following. In the Kraus representation, given Kraus operators $\{A_k\}_k$ of channel $\map A$ and $\{B_l\}_l$ of $\map B$, the Kraus operators of $S_\switch(\map A, \map B)$ are $\{\ketbra{0}{0} \otimes B_l A_k + \ketbra{1}{1} \otimes A_k B_l\}_{kl}$. As mentioned above, there exists a deterministic and exact simulation of the quantum switch with a single query to a general channel $\map B$ and two queries to a unitary channel $\map A$. Simulating the quantum switch for general $\map A$ and $\map B$ requires correlating each Kraus operator $A_k$ on the $\ket{0}$ branch---which can be obtained by querying $\map A$ before $\map B$---with the same $A_k$ on the $\ket{1}$ branch---which can be obtained by querying $\map A$ after $\map B$. In this view, $\map B$ can be considered as a fixed channel \cite{guerin2018observer}, and therefore the intuition is that querying it multiple times is no better than querying it once. The rational for the bound to be $\Theta(2^n)$ is that all the main steps in the proof of Theorem \ref{theo:main} except one (i.e., Lemmas \ref{lem:linearity_MN}, \ref{lem:p_as_vector_MN} and \ref{lem:qccc_invalid}) hold for a bound of $\Theta(2^n)$ with $(M+N)$-slot supermaps, and only for Lemma \ref{lem:indep_M1} were we only able to prove the $(M+1)$-slot case.

Another open question is whether or not a deterministic and exact simulation 
of the quantum switch is achievable by a fixed-order or classically-controlled-order quantum circuit with finitely many $M,N > \max(2, 2^n-1)$ slots. 
This question is similar in spirit to the question of performing a deterministic and exact transformation of a black-box unitary operator $\map U$, such as inversion, transposition, conjugation, or controlization 
\cite{araujo2014quantum,chiribella2016optimal,bisio2016quantum,sardharwalla2016universal,soleimanifar2016no,navascues2018resetting,chiribella2019quantum,dong2019controlled,sedlak2019optimal,quintino2019probabilistic,quintino2019reversing,miyazaki2019complex,trillo2020translating,quintino2022deterministic,grinko2022linear,ebler2023optimal,trillo2023universal,schiansky2023demonstration,yoshida2023reversing,mo2024parameterized,chen2024quantum,odake2024analytical}. For the case of unitary inversion, recent work has shown that at least $\Omega(4^n)$ queries to the unitary are needed \cite{odake2024analytical} and, conversely, that this bound is achievable \cite{yoshida2023reversing, mo2024parameterized, chen2024quantum}.  
On the other hand, unitary controlization can never be done exactly (even probabilistically) with a finite number of copies~\cite{gavorova2024topological}. It remains to be seen whether the action of the quantum switch can also be simulated with a finite number of queries to one or more of the channels. 

In this work, we have focused on deterministic and exact simulation. In practice, however, one might  be satisfied with a deterministic \textit{approximate} simulation---with some approximation parameter $\epsilon$---or in a \textit{probabilistic} exact simulation with some success probability $p$. In a companion paper \cite{bavaresco2024switch}, we study such questions using the techniques of semidefinite programming, where we present explicit upper bounds on the maximum success probability in the scenario where $n=1$ and the simulation is made with arbitrary four-slot combs.

{\it Conclusions.---}In this Letter, we have shown that 
the (one-sided) quantum query complexity of the action of the quantum switch, with respect to all supermaps with fixed or classically controlled causal order, is lower bounded by $ \max(3, 2^n)$.
This demonstrates an exponential separation in quantum query complexity between higher-order quantum transformations with indefinite causal order and standard quantum circuits, as a function of the number of qubits. Notably, the separation that we prove is formulated with respect to a computational task where the inputs and outputs of the computation are given by black-box quantum channels \cite{chiribella2008quantum,chiribella2008transforming,chiribella2013quantum,chiribella2018indefinite}. This is in contrast to previous works on the  query complexity of the quantum switch, where the output of the computation is a bit representing the evaluation of a classical function, in which case no such exponential separation has been found \cite{chiribella2012perfect,araujo2014computational,renner2022computational,abbott2023quantum,taddei2021computational}. Our work opens up the study of  query complexity in the context of higher-order quantum computation, where the inputs and outputs of the computation are quantum channels.

\begin{acknowledgments}
\textit{Acknowledgments.---}We would like to thank Alastair Abbott, Emanuel-Cristian Boghiu, Cyril Branciard, Anne Broadbent, Giulio Chiribella, Arthur Mehta, Simon Milz, Pierre Pocreau, Louis Salvail, Mykola Semenyakin, Jacopo Surace and Matt Wilson for helpful discussions. 
This work was supported by the MEXT Quantum Leap Flagship Program (MEXT QLEAP)  JPMXS0118069605, JPMXS0120351339, the Japan Society for the Promotion of Science (JSPS) KAKENHI Grant Number 21H03394 and 23K21643, and IBM Quantum.
S.Y.\ acknowledges support by JSPS KAKENHI Grant Number 23KJ0734, FoPM, WINGS Program, the University of Tokyo, and DAIKIN Fellowship Program, the University of Tokyo.
P.T.\ acknowledges funding from the JSPS Postdoctoral Fellowships for Research in Japan. J.B.\ acknowledges funding from the Swiss National Science Foundation (SNSF) through the funding schemes SPF and NCCR SwissMAP. 
Research at Perimeter Institute is supported in part by the Government of Canada through the Department of Innovation, Science and Economic Development and by the Province of Ontario through the Ministry of Colleges and Universities.

\end{acknowledgments}

%\bibliography{references.bib}


\clearpage
\onecolumngrid
\appendix

\renewcommand{\theequation}{S\arabic{equation}}
\renewcommand{\thetable}{S\arabic{table}}
\renewcommand{\thefigure}{S\arabic{figure}}
\renewcommand{\thetheorem}{S\arabic{theorem}}
\setcounter{equation}{0}
\setcounter{table}{0}
\setcounter{figure}{0}
\setcounter{theorem}{0}

\begin{center}
    \textbf{\large Supplemental Material for: Exponential separation in quantum query complexity of the quantum switch with respect to simulations with standard quantum circuits}
\end{center}

\FloatBarrier

\section{Problem Setting}

\begin{figure}[H]
    \centering
    \includegraphics[width=.8\linewidth]{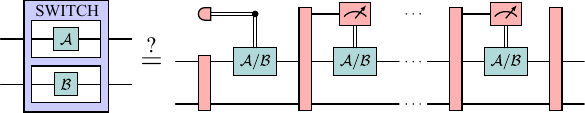}
    \caption{This work considers the question of simulating the action of the quantum switch of two black-box quantum channels $\mcA$ and $\mcB$ (\textit{left}), for all $\mathcal{A}$ and $\mathcal{B}$,  using a quantum circuit with classical control of the causal order (QC-CC) (\textit{right}).
    In the QC-CC shown on the right, a black box $\mcA$ or $\mcB$ is called depending on the previous  measurement outcome;  $\mcA$ and $\mcB$ are called $M$ and $N$ times in total, respectively. 
    }
    \label{fig:qccc_simulation}
\end{figure}

\section{Proof of Theorem \ref{theo:main}}

\renewcommand\thetheorem{\thesection\arabic{theorem}} 

\begin{theorem}[expanded]
Let $\map S_\switch :  [\mbL(I) \to \mbL(O)] \otimes  [\mbL(I') \to \mbL(O')]  \to [\mbL(P_C \otimes P_T) \to \mbL(F_C \otimes F_T)]$ be the quantum switch supermap, where $I, O, I', O', P_T, F_T, $ correspond to  $n$-qubit Hilbert spaces and $P_C, F_C$ correspond to qubit Hilbert spaces. Then, there is no ($M+1$)-slot supermap $ \map C : \bigotimes_{i=1}^{M}[\mbL(I_i) \to \mbL(O_i)] \otimes [\mbL(I'_1) \to \mbL(O'_1)] \to [\mbL(P_C \otimes P_T) \to \mbL(F_C \otimes F_T)]$, where $\{I_i\}_{i}, \{O_i\}_{i}, I'_1, O'_1$ correspond to  $n$-qubit Hilbert spaces, with fixed causal order or classical control of the causal order satisfying
    \begin{align}
        \supermap{C}( \underbrace{\map{A},\dots, \map{A}}_M, \map{B} ) &= \map S_\switch(\map{A}, \map{B})\label{eq:switch_simulation_1}
    \end{align}
    for all mixed unitary channels $\map{A}$ and unitary channels $\map{B}$, if $M \leq  \max(2, 2^{n}-1)$.
\end{theorem}

%\main*

\begin{proof} The proof is based upon a series of lemmas, proven below. First, assume that there exists a supermap $\map C$ such that Eq.\ \eqref{eq:switch_simulation_1} holds. Since $\mathcal{C}$ is a QC-CC supermap, we invoke Lemma \ref{lem:linearity_MN} with $N=1$, showing that the Choi operator $C$ of $\map C$ satisfies 
    \begin{align}
    	&C= \sum_{\vec{r}_{M+1} \in \mathrm{Perm}(1, \dots, M+1)}  	C_{P \vec{r}_{M+1} F}  \\ 
        &\textup{where}\quad\quad
       	C_{P \vec{r}_{M+1} F} = \sum_{a} \dketbra{C_{P \vec{r}_{M+1} F}^{(a)}} \quad \forall \, \vec{r}_{M+1}  \, ,
\end{align}
such that for every $a$ and for every $\vec{r}_{M+1} \in \mathrm{Perm}(1, \dots, M+1)$ (i.e.,\ a vector representing a permutation of the integers from 1 to $M+1$), we have 
\begin{align}
      \dket{C_{P \vec{r}_{M+1} F}^{(a)}}  \ast   \bigotimes_{i=1}^{M} \dket{U_i}
         \otimes \dket{V_1}
          = \sum_{k}^{M} 
          \xi^{(a), \vec{r}_{M+1}}_{k1}(\{U_i\}_i,\ V_1)
          \dket{S_\switch}\ast (\dket{U_k}\otimes \dket{V_1}) \, , \label{eq:pure_C_ar_M1}
\end{align}
for some $\{ \xi^{(a),\vec{r}_{M+1}}_{k1}(\{U_i\}_i,\ V_1) \}_{k} \in \mbC$.
    
Then, by Lemma \ref{lem:indep_M1}, taking $n$ and $M$ such that $M<4^n/2 +2$ [which is implied by $M\leq \max(2,2^n-1)$], 
there exists a set of complex numbers $\{ \tilde{\xi}^{(a), \vec{r}_{M+1}}_{k1} \}_k$, such that Eq.\ \eqref{eq:pure_C_ar_M1} is also satisfied for the reassignment $\xi^{(a), \vec{r}_{M+1}}_{k1} \leftarrow \tilde{\xi}^{(a), \vec{r}_{M+1}}_{k1} $ such that for all $k \in \{1, \dots, M\}$, $\tilde{\xi}^{(a), \vec{r}_{M+1}}_{k1}$ is simultaneously
\begin{enumerate}
\item  independent of $U_{k}$ and $V_1$, and 
\item  linear in $U_{k'}$ for all $k' \neq k$.
\end{enumerate}

Invoking Lemma \ref{lem:p_as_vector_MN} with $N=1$, this then implies that
    \begin{align} \label{eq:pure_C_lemmapasvector_thm1}
      &\dket{C_{P \vec{r}_{M+1} F}^{(a)}}^{PF I_1 O_1 \cdots  I_M O_M  I'_1 O'_1 }
= \sum_{k=1}^M 
\dket{S_\switch}^{PF I_k O_k I'_1 O'_1} \otimes
        \dket{\tilde{\xi}^{(a),\vec{r}_{M+1}}_{k1}}^{\{I_1 O_1, \cdots , I_M O_M \} \setminus \{I_k O_k\} } 
    \end{align}
    for some vectors $\dket{\tilde{\xi}^{(a),\vec{r}_{M+1}}_{k1}}^{\{I_1 O_1, \cdots , I_M O_M \} \setminus \{I_k O_k\} }$.
Therefore, for all $\vec{r}_{M+1} \in \mathrm{Perm}(1, \dots, M+1)$, we have
\begin{align} \label{eq:final_decomp}
       	C_{P \vec{r}_{M+1} F}
        = \sum_{a} \dketbra{C_{P \vec{r}_{M+1} F}^{(a)}} \, ,
\end{align}
with each $\dket{C_{P \vec{r}_{M+1} F}^{(a)}}$ defined by Eq.\ \eqref{eq:pure_C_lemmapasvector_thm1}.

Finally, from Lemma \ref{lem:qccc_invalid}, we find that a supermap with Choi operator $C= \sum_{\vec{r}_{M+1} \in \mathrm{Perm}(1, \dots, M+1)}  	C_{P \vec{r}_{M+1} F}$, with $	C_{P \vec{r}_{M+1} F}$ satisfying Eqs.\ \eqref{eq:pure_C_lemmapasvector_thm1} and \eqref{eq:final_decomp} for all $\vec{r}_{M+1} \in {\rm Perm}(1 ,\dots ,M+1 )$ cannot have fixed causal order or classical control of the causal order if $M \leq \max(2, 2^n-1)$. 
\end{proof}

\setcounter{theorem}{0}

\vspace{3ex}
%%%%%%%%%%%%%%%%%%%%%%%%%%%%%%%%%%%%%%%%%%%%%%%%%

\begin{lemma}\label{lem:span}
    Let $\ket{\phi}$ and $\{\ket{\psi_i}\}_i$ be vectors in $\mathbb{C}^d$. 
\begin{align}
      \text{If }
\ketbra{\phi}\leq \sum_i \ketbra{\psi_i}, \;
\text {then} \ket{\phi}\in\mathrm{span}(\{\ket{\psi_i}\}).
\end{align}    
That is, there exist complex numbers $\alpha_i$ such that $\ket{\phi}=\sum_i \alpha_i \ket{\psi_i}$.
\end{lemma}

\begin{proof}
The proof will go by contradiction. We start by pointing out that any vector $\ket{\phi}\in\mathbb{C}^d$ can be decomposed as 
\begin{align}
    \ket{\phi} = \ket{\psi}+\ket{\psi_\perp},
\end{align}
where $\ket{\psi}\in \mathrm{span}(\{\ket{\psi_i}\})$, $\ket{\psi_\perp}\notin \mathrm{span}(\{\ket{\psi_i}\})$. Also, since $\ket{\psi_\perp}\notin \mathrm{span}(\{\ket{\psi_i}\})$, we have that $\braket{\psi_\perp}{\psi_i}=0$ for every $i$.

Now, assume that  $\ket{\phi}\notin\mathrm{span}(\{\ket{\psi_i}\})$. In this case, we necessarily have that $\ket{\psi_\perp}\neq0$.
Using this decomposition $\ket{\phi} = \ket{\psi}+\ket{\psi_\perp}$, we can write the inequality $\ketbra{\phi}\leq \sum_i \ketbra{\psi_i},$ as 
\begin{align} \label{eq:psi_psiperp}
    \ketbra{\psi} + \ketbra{\psi}{\psi_\perp} +  \ketbra{\psi_\perp}{\psi} + \ketbra{\psi_\perp} \leq \sum_i \ketbra{\psi_i}.
\end{align}
We then apply $\bra{\psi_\perp}$ and  $\ket{\psi_\perp}$ on both sides of operator inequality~\eqref{eq:psi_psiperp} to obtain the real number inequality
\begin{align} 
    \braket{\psi_\perp} \braket{\psi_\perp} \leq 0.
\end{align}
However, since $\ket{\psi_\perp}\neq0$, $\braket{\psi_\perp} \braket{\psi_\perp}$ is strictly positive, hence we have arrived at a contradiction. Therefore, $\ket{\phi}$ must belong to the $\mathrm{span}(\{\ket{\psi_i}\})$.
\end{proof}

\vspace{3ex}
%%%%%%%%%%%%%%%%%%%%%%%%%%%%%%%%%%%%%%%%%%%%%
\begin{lemma}\label{lem:linearity_MN}
Let $\map S_\switch :  [\mbL(I) \to \mbL(O)] \otimes  [\mbL(I') \to \mbL(O')]  \to [\mbL(P_C \otimes P_T) \to \mbL(F_C \otimes F_T)]$ be the quantum switch supermap, where $I, O, I', O', P_T, F_T, $ correspond to  $n$-qubit Hilbert spaces and $P_C, F_C$ correspond to  qubit Hilbert spaces. Then, if there exists an $(M+N)$-slot supermap $ \map C : \bigotimes_{i=1}^{M}[\mbL(I_i) \to \mbL(O_i)] \otimes \bigotimes_{j=1}^{N}[\mbL(I'_j) \to \mbL(O'_j)] \to [\mbL(P_C \otimes P_T) \to \mbL(F_C \otimes F_T)]$, where $\{I_i\}_{i}, \{O_i\}_{i}, \{I'_j\}_{j}, \{O'_j\}_{j}$ correspond to  $n$-qubit Hilbert spaces, with fixed causal order or classical control of causal order, satisfying
    \begin{align}
        \supermap{C}( \underbrace{\map{A},\dots, \map{A}}_M, \underbrace{\map{B}, \dots , \map{B}}_N ) &= \map S_\switch(\map{A}, \map{B})\label{eq:switch_simulation_2}
    \end{align}
    for all mixed unitary channels $\map{A}$ and $\map{B}$ (or, if $N=1$, for all mixed unitary channels $\map{A}$ and unitary channels $\map{B}$), then the Choi operator $C$ of $\map C$ satisfies
    \begin{align}
    	&C= \sum_{\vec{r}_{M+N} \in \mathrm{Perm}(1, \dots, M+N)}  	C_{P \vec{r}_{M+N} F}  \\
    	&\textup{where}\quad\quad
       	C_{P \vec{r}_{M+N} F} = \sum_a \dketbra{C_{P \vec{r}_{M+N} F}^{(a)}} \quad \forall \, \vec{r}_{M+N} \, ,
\end{align}
such that for every $a$ and for every $\vec{r}_{M+N} \in \mathrm{Perm}(1, \dots, M+N)$ (i.e.,\ a vector representing a permutation of the integers from 1 to $M+N$), 
    \begin{align}
      \dket{C_{P \vec{r}_{M+N} F}^{(a)}}  \ast   \bigotimes_{i=1}^{M} \dket{U_i}
          \bigotimes_{j=1}^{N} \dket{V_j}
          = \sum_{k=1}^{M} \sum_{l=1}^{N}
          \xi^{(a),\vec{r}_{M+N}}_{kl}(\{U_i\}_i,\ \{V_j\}_j)
          \dket{S_\switch}\ast (\dket{U_k}\otimes \dket{V_l}) \, , \label{eq:pure_C_ar}
    \end{align}
    for some $\{\xi^{(a),\vec{r}_{M+N}}_{kl}(\{U_i\}_i,\ \{V_j\}_j) \}_{kl} \in \mbC$.
\end{lemma}

\begin{proof}
Assume that Eq.\ \eqref{eq:switch_simulation_2} holds for all mixed unitary channels $\map A, \map B$, for some integers $M,N \ge 1$ and some given qubit number $n$. Then, for any sets of unitary channels $\{\map A_1, \dots \map A_K\}$ and $\{\map B_1, \dots \map{B}_L\}$ with $K,L \ge 1 $, we have
\begin{equation}
	\map C \left(
	 \sum_{i_1 = 1}^K  \frac{\map A_{i_1}}K , \dots,  \sum_{i_M = 1}^K  \frac{\map A_{i_M}}K ;  
	 \sum_{j_1 = 1}^L \frac{\map B_{j_1}}L , \dots, \sum_{j_N = 1}^L \frac{\map B_{j_N}}L
	\right)
	=
	\map S_\switch
	\left(
	\sum_{p = 1}^K \frac{\map A_{p}}K ,
	\sum_{q= 1}^L \frac{\map B_{q}}L
	\right) \, .
\end{equation}
(For $N=1$, it is sufficient to assume that Eq.\ \eqref{eq:switch_simulation_2} holds for all mixed unitary channels $\map{A}$ and unitary channels $\map{B}$, in which case we take $L=1$.)
By the multilinearity of $\map C$ and $\map S_\switch$, this then implies that
\begin{align}
\frac{1}{K^M L^N} \sum_{i_1, \dots, i_M = 1}^K \; \sum_{ j_1, \dots , j_N = 1}^L
		\map C \left( 
  \map A_{i_1}, \dots,   \map A_{i_M} ;  
 \map B_{j_1}, \dots, \map B_{j_N}
	\right)
	=
	\frac{1}{KL}
		\sum_{p=1}^K \sum_{q=1}^L
	\map S_\switch
	\left(
\map A_{p},
 \map B_{q}
	\right) \, .
\end{align}
Rewriting this expression in the Choi representation -- using the convention that $C, S_{\switch}, A_i, B_j$ are the Choi matrices of  $\map C, \map S_{\switch}, \map{A}_i, \map{B}_j$, respectively -- gives
\begin{align}
	\frac{1}{K^M L^N} 	\sum_{i_1, \dots, i_M }^K \sum_{ j_1, \dots , j_N = 1}^L
	 C \ast \left(
	 	 A_{i_1} \otimes \dots  \otimes     A_{i_M} \otimes   
	 B_{j_1}  \otimes  \dots  \otimes   B_{j_N}
	\right)
	=
	\frac{K^{M-1} L^{N-1}}{K^M L^N}
	\sum_{p=1}^K \sum_{q=1}^L
	 S_\switch \ast
	\left(
	 A_{p} \otimes
	 B_{q}
	\right) \, .
\end{align}
$K L$ number of terms on the left-hand side can be written using Eq.\ \eqref{eq:switch_simulation_2} in the form of the quantum switch, leading to the equation
\begin{align}
	C \ast \frac{1}{K^M L^N} 	\sum_{i_1, \dots, i_M =1 }^K  \sum_{\substack{j_1, \dots , j_M = 1 \\ \neg (i_1=\dots = i_M \land j_1 = \dots = j_M)}}^L
	 \left(
	\bigotimes_{k=1}^M  A_{i_k} \bigotimes_{l=1}^N  B_{j_l} 
	\right)
	=
 S_\switch \ast
	\frac{K^{M-1} L^{N-1} -1 }{K^M L^N}
	\sum_{p=1}^K \sum_{q=1}^L
	\left(
	 A_{p} \otimes
	 B_{q}
	\right) \, .
\end{align}	
Now, assuming that $\map C$ is a quantum supermap with fixed causal order or with classical control of the causal order, then its Choi matrix $C$ satisfies the following relation \cite{wechs2021quantum}
   \begin{align}
	&C= \sum_{\vec{r}_{M+N} \in \mathrm{Perm}(1, \dots, M+N)}  	C_{P \vec{r}_{M+N} F}  \\
	&\textup{where}\quad\quad
	C_{P \vec{r}_{M+N} F} \ge 0 \quad \forall \, \vec{r}_{M+N} \, . \label{eq:choi_positivity}
\end{align}
From Eq.~(\ref{eq:choi_positivity}), $C_{P \vec{r}_{M+N} F}$ can be diagonalized as
\begin{align}
	C_{P \vec{r}_{M+N} F} = \sum_a \dketbra{C_{P \vec{r}_{M+N} F}^{(a)}} \,.
\end{align}
Note that we can also write $S_\switch = \dketbra{S_\switch}$, where 
\begin{align}
	\dket{S_\switch}^{PFIO I' O'}
 &\coloneqq \ket{0}^{P_C}\ket{0}^{F_C}\dket{\1}^{P_T I}\dket{\1}^{O I'}\dket{\1}^{O'F_T} + \ket{1}^{P_C}\ket{1}^{F_C}\dket{\1}^{P_T I'}\dket{\1}^{O'I}\dket{\1}^{O F_T} \, ,
\end{align}
with $P$ and $F$ corresponding to  joint Hilbert spaces defined by $P\coloneqq P_C\otimes P_T$ and $F\coloneqq F_C\otimes F_T$. The above equations together imply that
    \begin{align}\label{eq:big_sum}
        0 &\leq \sum_a \dketbra{C_{P \vec{r}_{M+N} F}^{(a)}}  \ast 
    	\sum_{i_1, \dots, i_M =1 }^K  \sum_{\substack{j_1, \dots , j_M = 1 \\ \neg (i_1=\dots = i_M \land j_1 = \dots = j_M)}}^L
	 \left(
	\bigotimes_{k=1}^M  A_{i_k} \bigotimes_{l=1}^N B_{j_l} 
	\right)
	\\
        &\leq \dketbra{S_\switch}\ast
	[K^{M-1} L^{N-1} -1  ]
	\sum_{p =1}^K \sum_{q=1}^L
	\left(
	 A_{p} \otimes
	 B_{q}
	\right)
    \end{align}
for all $\vec{r}_{M+N} \in {\rm Perm}(1 \dots, M+N )$. 
    
Consider now the case where $K=M, L=N$.
Since the channels are given by unitary channels, the Choi operators are given by $A_i = \dketbra{U_i}$ and $B_j = \dketbra{V_j}$. Since the right-hand side of Eq.\ \eqref{eq:big_sum} is a sum of positive operators, the inequality also holds for the sum of any subset of the terms on the right-hand side.  Then, we obtain (by considering only the term in the  sums over $i_1, \dots, i_M, j_1, \dots, j_N$ corresponding to $i_k=k, j_l = l, \forall \, k,l$) that 
    \begin{align}
          0 &\leq \sum_a \dketbra{C_{P \vec{r}_{M+N} F}^{(a)}}  \ast 
          \bigotimes_{i=1}^M \dketbra{U_i}
          \bigotimes_{j=1}^N \dketbra{V_j}
          \\ &\leq 
          \dketbra{S_\switch}\ast 
          	[M^{(M-1)} N^{(N-1)} - 1 ]
	\sum_{k=1}^M 	\sum_{l=1}^N
	\left[
	\dketbra{U_k} \otimes \dketbra{V_l}
	\right] \, ,
    \end{align}
    for all $\vec{r}_{M+N} \in {\rm Perm}(1 \dots, M+N )$. Therefore, for every $a$ and for every $\vec{r}_{M+N} \in {\rm Perm}(1 \dots, M+N )$, we have that
    \begin{align}
      \dket{C_{P \vec{r}_{M+N} F}^{(a)}}  \ast   \bigotimes_{i=1}^M \dket{U_i}
          \bigotimes_{j=1}^N \dket{V_j}
          = 	\sum_{k=1}^M 	\sum_{l=1}^N
         \xi^{(a),\vec{r}_{M+N}}_{kl}(\{U_i\}_i,\ \{V_j\}_j)
          \dket{S_\switch}\ast (\dket{U_k}\otimes \dket{V_l}) \, , 
    \end{align}
    for some $\{\xi^{(a),\vec{r}_{M+N}}_{kl}(\{U_i\}_i,\ \{V_j\}_j) \}_{kl} \in \mbC$.
\end{proof}

\vspace{3ex}
%%%%%%%%%%%%%%%%%%%%%%%%%%%%%%%%%%%%%%%%%%%%%%%%%
\begin{lemma}\label{lem:indep_M1}
Let $C \in \mbL(I_1 \otimes \cdots \otimes I_M \otimes O_1 \otimes \cdots \otimes O_M 
\otimes I'_1 \otimes O'_1 \otimes P_C \otimes P_T \otimes F_C \otimes F_T) $, for some $M \in \mathbb{N}^+$ where $P_T, F_T, \{I_i\}_{i}, \{O_i\}_{i}, I'_1, O'_1$ correspond to  $n$-qubit Hilbert spaces for some $n \in \mathbb{N}^+$, and $P_C, F_C$ correspond to  qubit Hilbert spaces, be a linear operator such that $C = \dketbra{C}{C}$ for some vector $\dket{C}$. If, for all $(M+1)$-tuples of $n$-qubit unitary operators $(U_1, \dots, U_M, V_1)$,
\begin{equation}\label{eq:indep_M1}
    \dket{C} * \dket{U_1}^{I_1 O_1} \otimes \dots \otimes \dket{U_M}^{I_M O_M} \otimes  \dket{V_1}^{I'_1 O'_1} = \sum_{k=1}^M \xi_{k1} \dket{S_\switch} * \dket{U_k}^{I_k O_k} \otimes \dket{V_1}^{I'_1 O'_1}
\end{equation}
for some complex numbers $\xi_{k1} := \xi_{k1}(\{U_i\}_i,\ V_1) \in \mathbb{C}$, then there exist complex numbers $\tilde{\xi}_{k1} := \tilde{\xi}_{k1} (\{U_i\}_i,\ V_1) \in \mathbb{C} $ such that Eq.\ \eqref{eq:indep_M1} with  $\{{\xi}_{k1}\}_{k=1}^M \leftarrow \{\tilde{\xi}_{k1}\}_{k=1}^M$ remains satisfied and, for all $k \in \{1, \dots, M\}$, $\tilde{\xi}_{k1}$ is simultaneously
\begin{enumerate}
    \item  independent of $U_{k}$ and $V_1$ (independence condition), and 
    \item  linear in $U_{i}$ for all $i \neq k$ (linearity condition),
\end{enumerate}
as long as $M < 4^{n}/2 +2$. 
\end{lemma}

\begin{proof}
Assume that Eq. \eqref{eq:indep_M1} holds. Then, in particular, it holds for the choice $U_i=\sigma_{\vec{r_i}}$ for $i\in \{1,\ldots,M\}$ and $V_1=\sigma_{\vec{q}_1}$, where $\vec{r}_i,\vec{q}_1\in \{0,1,2,3\}^{\times n}$ and $\{\sigma_{\vec{r}_i}\}_i,\ \sigma_{\vec{q}_1}$ are $n$-qubit Pauli operators.
Here, the set of $n$-qubit Pauli operators is defined by
\begin{align}
    \left\{\sigma_{\vec{r}}\coloneqq \bigotimes_{i=1}^{n} \sigma_{r_i}\middle|\vec{r}\in\{0,1,2,3\}^{\times n}\right\},
\end{align}
where $\sigma_0, \sigma_1, \sigma_2, \sigma_3$ are 1-qubit Pauli operators defined by
\begin{align}
    \sigma_0 :=\left(\begin{matrix}
        1 & 0\\
        0 & 1
    \end{matrix}
    \right),\quad 
    \sigma_1\coloneqq \left(\begin{matrix}
        0 & 1\\
        1 & 0
    \end{matrix}
    \right),\quad
    \sigma_2\coloneqq
    \left(\begin{matrix}
        0 & -i\\
        i & 0
    \end{matrix}
    \right),\quad
    \sigma_3\coloneqq \left(\begin{matrix}
        1 & 0\\
        0 & -1
    \end{matrix}
    \right).
\end{align}
Thus,
\begin{equation}
    \dket{C} * \dket{\sigma_{\vec{r}_1}}^{I_1 O_1} \otimes \dots \otimes \dket{\sigma_{\vec{r}_M}}^{I_M O_M} \otimes  \dket{\sigma_{\vec{q}_1}}^{I'_1 O'_1} = \sum_{k=1}^M \xi_{k1} \dket{S_\switch} * \dket{\sigma_{\vec{r}_k}}^{I_k O_k} \otimes \dket{\sigma_{\vec{q}_1}}^{I'_1 O'_1}.
\end{equation}
Now suppose that $F \ge 1$ of the $\{\sigma_{\vec{r}_i} \}_{i=1}^M$ are equal to some fixed $n$-qubit Pauli operator $\sigma_{\vec{w}}$. Let the set of integers labelling those Pauli operators be denoted $\mathbb{F}:= \{1 \leq i \leq M | \sigma_{\vec{r}_i}=\sigma_{\vec{w}} \} $, such that $|\mathbb{F}|=F$. Equation \eqref{eq:indep_M1} then reads 
\begin{equation}\label{eq:indep_w_M1}
\begin{split}
 \dket{C} * \bigotimes_{i \in \mathbb{F}} \dket{\sigma_{\vec{w}}}^{I_i O_i} \otimes \bigotimes_{i \in \{1,\dots , M\}\backslash \mathbb{F}}
  \dket{\sigma_{\vec{r}_{i}}}^{I_{i} O_{i}} 
 \otimes \dket{\sigma_{\vec{q}_1}}^{I'_1 O'_1} 
 = \sum_{k=1}^M \xi_{k1}^{\{\vec{r}_l\}_l,\vec{q}_1}  \dket{S_\switch} * \dket{\sigma_{\vec{r}_k}}^{I_k O_k} \otimes \dket{\sigma_{\vec{q}_1}}^{I'_1 O'_1} \, ,
\end{split}
\end{equation}
where, for the input unitaries chosen as $n$-qubit Paulis, we define
\begin{align}
\xi_{k1}^{\{\vec{r}_l\}_l,\vec{q}_1} 
   :=
\xi_{k1}(  \{U_i=\sigma_{\vec{r}_i}\}_{i =1}^M ,V_1=\sigma_{\vec{q}_1}  )\add{.}
\end{align} 
In the following, we will adopt the shorthand convention that any changes to the dependence of $\xi_{k1}$ from $\xi_{k1}^{\{\vec{r}_l\}_l,\vec{q}_1}$ will be specified as $\xi_{k1}^{\{\vec{r}_l\}_l,\vec{q}_1}[U_i= (\dots), V_j = (\dots)]$, with all unspecified arguments $U_i, V_1$ defined to be the same as for $\xi_{k1}^{\{\vec{r}_l\}_l,\vec{q}_1}$ defined above.
A key point that we note for later is that the value of each individual variable $\xi_{k1}^{\{\vec{r}_l\}_l,\vec{q}_1}$ for $k \in \mathbb{F}$ is not uniquely determined from Eq.~\eqref{eq:indep_M1}, so we can take a different set of variables $\tilde{\xi}_{k1}$ still satisfying Eq.~\eqref{eq:indep_M1}.

We now show that for all $k \in \{1,\dots, M\}$, the variables $\xi_{k1}$ can be 
replaced by $\tilde{\xi}_{k1}$ defined by
\begin{align}
\tilde{\xi}_{k1}(\{U_i=\sigma_{\vec{r}_i}\}_{i=1}^M,V_1=\sigma_{\vec{q}_1})&:=\xi_{k1}^{\{\vec{r}_l\}_l,\vec{q}_1}[U_k=\sigma_{\vec{r}^*_k}, V_1=\sigma_{\vec{q}^*}] \, ,
    \label{eq:global_def_M1}
    \\
    \tilde{\xi}_{k1}\left(\left\{U_i=\sum_{\vec{r}_i}{\alpha_{\vec{r}_i}^i \sigma_{\vec{r}_i}}\right\}_{i=1}^M,\ V_1=\sum_{\vec{q}_1}\beta_{\vec{q}_1}^1\sigma_{\vec{q}_1}\right)&:=
    \sum_{\{\vec{r}_i\}_{i\neq k}}
    \left(
    \prod_{i\neq k}
    \alpha^i_{\vec{r}_i}\right)
    \tilde{\xi}_{k1}(\{U_i=\sigma_{\vec{r}_i}\}_{i=1}^M,V_1=\sigma_{\vec{q}_1}) \, ,
    \label{eq:global_def_M1_linear}
\end{align}
where $\mathbb{F}_k:=\{1\leq i\leq M\mid \sigma_{\vec{r}_i}=\sigma_{\vec{r}_k}\}$, $\vec{r}^*_k \in \{0,1,2,3\}^{\times n}$ is an arbitrary vector \textit{outside} of the set $\{\vec{r}_1,\ldots,\vec{r}_{k-1}, \vec{r}_{k+1},\ldots,\vec{r}_M\}$, $\vec{q}^* \in \{0,1,2,3\}^{\times n}$ is an arbitrary fixed vector, and $\alpha_{\vec{r}_i}^i, \beta_{\vec{q}_1}^1$ are complex numbers. 
In the discussion below, we pick one choice of $\vec{r}_k^*$ defined as a function of $\vec{r}_1,\ldots,\vec{r}_{k-1}, \vec{r}_{k+1},\ldots,\vec{r}_M$, i.e., $\vec{r}_k^* = \vec{r}_k^*(\vec{r}_1,\ldots,\vec{r}_{k-1}, \vec{r}_{k+1},\ldots,\vec{r}_M)$. Such an $\vec{r}_k^*$ always exists for $M < 4^n$.

By construction, if the definition in Eqs.\ \eqref{eq:global_def_M1}--\eqref{eq:global_def_M1_linear}
%is well-defined and 
satisfies Eq.\ \eqref{eq:indep_M1}, then $\tilde{\xi}_{k1}$ satisfies both the linearity and independence conditions outlined in the statement of the lemma. We now proceed to show that the definition in Eqs.\ \eqref{eq:global_def_M1}--\eqref{eq:global_def_M1_linear} indeed satisfies Eq.\ \eqref{eq:indep_M1}. We do this by considering the dependence of $\xi_{k1}^{\{\vec{r}_l\}_l,\vec{q}_1}$ on the unitaries $\{U_i=\sigma_{\vec{r}_i}\}_{i =1}^M $ and $V_1=\sigma_{\vec{q}_1}$  in turn.
\vspace{2ex}

\subsection{Dependence on $\{U_i\}_{i=1}^M$}

We focus on the case where $\{U_i\}_i,\ V_1$ are chosen from the set of Pauli operators. 
For every $k \in \{1, \dots M\}$, we choose one $\sigma_{\vec{r}^*_k}$ and take $\sigma$ to be the unique $n$-qubit Pauli operator such that $\sigma  \sigma_{\vec{r}_k} = \beta \sigma_{\vec{r}^*_k}$, where $\beta \in \{-1,1,i,-i\}$. We then consider the following expression
\begin{equation}
 \mathbf{E}:=  
 \frac{\d}{\d \theta} \bigg |_{\theta=0} 
 \left[
 \dket{C} 
 * \bigotimes_{m \in \mathbb{F}_k}  \dket{e^{i \theta \sigma } \sigma_{\vec{r}_m}}^{I_m O_m} 
 * \bigotimes_{m=1 | m \notin \mathbb{F}_k}^M \dket{\sigma_{\vec{r}_{m}}}^{I_{m} O_{m}} 
 * \dket{\sigma_{\vec{q}_1}}^{I'_1 O'_1} 
 \right] \, .
\end{equation}
Due to linearity, this expression can be evaluated in two ways: either by first computing the derivative and then applying Eq.\ \eqref{eq:indep_M1}, or by first applying Eq.\ \eqref{eq:indep_M1} and then computing the derivative. The former method gives
\begin{align}
 \mathbf{E} &=  
 i \beta \sum_{m\in \mathbb{F}_k}
 \left[
 \dket{C} 
 * \dket{ \sigma_{\vec{r}^*_k}}^{I_m O_m} 
 * \bigotimes_{i \in \mathbb{F}_k | i \neq m}  \dket{ \sigma_{\vec{r}_{i}}}^{I_{i} O_{i}} 
 * \bigotimes_{i=1 | i \notin \mathbb{F}_k}^M \dket{\sigma_{\vec{r}_{i}}}^{I_{i} O_{i}} 
 *  \dket{\sigma_{\vec{q}_1}}^{I'_1 O'_1} 
 \right] \notag
 \\ \nonumber
 &= 
 i \beta \sum_{m\in \mathbb{F}_k} 
 \Bigg[
 \xi_{m1}^{\{\vec{r}_l\}_l,\vec{q}_1} [U_m = \sigma_{\vec{r}^*_k}]
 \dket{S_\switch} * \dket{\sigma_{\vec{r}^*_k}} * \dket{\sigma_{\vec{q}_1}} 
 \\
 &~~~~~~~~+
 \sum_{i \in \mathbb{F}_k | i \neq m} \xi_{i1}^{\{\vec{r}_l\}_l,\vec{q}_1} [U_m = \sigma_{\vec{r}^*_k}]
 \dket{S_\switch} * \dket{\sigma_{\vec{r}_k}} * \dket{\sigma_{\vec{q}_1}} \nonumber
 \\ 
 &~~~~~~~~+
 \sum_{i=1 | i \notin \mathbb{F}_k}^M \xi_{i1}^{\{\vec{r}_l\}_l,\vec{q}_1} [U_m = \sigma_{\vec{r}^*_k}]
 \dket{S_\switch} * \dket{\sigma_{\vec{r}_{i}}} * \dket{\sigma_{\vec{q}_1}}
 \Bigg] \, , 
\end{align}
while the latter gives
\begin{align}
	\mathbf{E} &=  
	\frac{\d}{\d \theta} \bigg |_{\theta=0} 
	\Bigg[
	 \sum_{m\in \mathbb{F}_k}
	\xi_{m1}^{\{\vec{r}_l\}_l,\vec{q}_1} 	 [\{ U_{i} = e^{i \theta \sigma} \sigma_{\vec{r}_k}\}_{i\in \mathbb{F}_k}]
	\dket{S_\switch} * \dket{e^{i \theta \sigma} \sigma_{\vec{r}_k}}^{I_m O_m} * \dket{\sigma_{\vec{q}_1}}^{I'_1 O'_1} 
	\notag \\
	&~~~~~~~~~~~~+
	 \sum_{m=1 | m\notin \mathbb{F}_k}^M \xi_{m1}^{\{\vec{r}_l\}_l,\vec{q}_1} [\{ U_{i} = e^{i \theta \sigma} \sigma_{\vec{r}_k}\}_{i\in \mathbb{F}_k}]
	\dket{S_\switch} * \dket{ \sigma_{\vec{r}_m}}^{I_m O_m} * \dket{\sigma_{\vec{q}_1}}^{I'_1 O'_1} 
	\Bigg] \notag
	\\
&=
i  \beta 
\sum_{m\in \mathbb{F}_k} 
\xi_{m1}^{\{\vec{r}_l\}_l,\vec{q}_1}
\dket{S_\switch} *  \dket{\sigma_{\vec{r}^*_k}} * \dket{\sigma_{\vec{q}_l}} \notag
\\ %\label{eq:line1_M1}
&~~~~~~~~~~~~+
	\frac{\d}{\d \theta} \bigg |_{\theta=0} 
\Bigg[
\sum_{m\in \mathbb{F}_k} 
\xi_{m1}^{\{\vec{r}_l\}_l,\vec{q}_1} [\{ U_{i} = e^{i \theta \sigma} \sigma_{\vec{r}_k}\}_{i\in \mathbb{F}_k}]
\Bigg]
\dket{S_\switch} * \dket{ \sigma_{\vec{r}_k}} * \dket{\sigma_{\vec{q}_1}} \notag
\\ %\label{eq:line2_M1}
&~~~~~~~~~~~~+
\sum_{\vec{v}\in \{\vec{r}_1,\ldots,\vec{r}_M\}\backslash\{\vec{r}_k\}}
	\frac{\d}{\d \theta} \bigg |_{\theta=0} 
\Bigg[
\sum_{m\in \mathbb{F}_{\vec{v}}}
\xi_{m1}^{\{\vec{r}_l\}_l,\vec{q}_1} [\{ U_{i} = e^{i \theta \sigma} \sigma_{\vec{r}_k}\}_{i\in \mathbb{F}_k}]
\Bigg] 
\dket{S_\switch} * \dket{ \sigma_{\vec{v}}}* \dket{\sigma_{\vec{q}_1}} 
\, ,
\end{align}
where $\mathbb{F}_{\vec{v}}:=\{1\leq m\leq M\mid \vec{r}_m=\vec{v}\}$. Note that the vector
\begin{align}
    \dket{ S_{\switch}}*\dket{\sigma_{\vec{r}_m}}^{I_m O_m}*\dket{\sigma_{\vec{q}_1}}^{I_1' O'_1}
\end{align}
belongs to a Hilbert space corresponding to  $P_T\otimes F_T$, which is independent of $I_m, O_m, I_1', O'_1$, thus the superscripts $I_m, O_m, I_1', O'_1$ can be omitted. Also, note that 
\begin{align}
    \sum_{m\in \mathbb{F}_k} 
 \xi_{m1}^{\{\vec{r}_l\}_l,\vec{q}_1} [\{ U_{i} = e^{i \theta \sigma} \sigma_{\vec{r}_k}\}_{i\in \mathbb{F}_k}]
\end{align}
and
\begin{align}
\sum_{m\in \mathbb{F}_{\vec{v}}}
\xi_{m1}^{\{\vec{r}_l\}_l,\vec{q}_1} [\{ U_{i} = e^{i \theta \sigma} \sigma_{\vec{r}_k}\}_{i\in \mathbb{F}_k}]
\end{align}
are differentiable since their values are uniquely determined from 
\begin{align}\label{eq:diff_M1}
    &\dket{C} 
 * \bigotimes_{m \in \mathbb{F}_k}  \dket{e^{i \theta \sigma } \sigma_{\vec{r}_m}}^{I_m O_m} 
 * \bigotimes_{m=1 | m \notin \mathbb{F}_k}^M \dket{\sigma_{\vec{r}_{m}}}^{I_{m} O_{m}} 
 * \dket{\sigma_{\vec{q}_1}}^{I'_1 O'_1} 
 \nonumber\\
 =&
 \left(
 \sum_{m\in \mathbb{F}_k} 
 \xi_{m1}^{\{\vec{r}_l\}_l,\vec{q}_1} [\{ U_{i} = e^{i \theta \sigma} \sigma_{\vec{r}_k}\}_{i\in \mathbb{F}_k}]
 \right)
 \dket{ S_{\switch}}*\dket{e^{i\theta \sigma}\sigma_{\vec{r}_k}}*\dket{\sigma_{\vec{q}_1}}
 \nonumber\\
 +&\sum_{\vec{v}\in \{\vec{r}_1,\ldots ,\vec{r}_M\}\backslash\{\vec{r}_k\}}
 \left(
\sum_{m \in\mathbb{F}_{\vec{v}}}
\xi_{m1}^{\{\vec{r}_l\}_l,\vec{q}_1} [\{ U_{i} = e^{i \theta \sigma} \sigma_{\vec{r}_k}\}_{i\in \mathbb{F}_k}]
 \right)
 \dket{S_\switch} * \dket{ \sigma_{\vec{v}}}* \dket{\sigma_{\vec{q}_1}} \, ,
\end{align}
and thus can be obtained from the inner product of the vector on the left-hand side of Eq.~(\ref{eq:diff_M1}) and vectors $\dket{ S_{\switch}}*\dket{e^{i\theta \sigma}\sigma_{\vec{r}_k}}*\dket{\sigma_{\vec{q}_1}}$ or $\dket{ S_{\switch}}*\dket{\sigma_{\vec{v}}}*\dket{\sigma_{\vec{q}_1}}$ (which are mutually orthogonal), which are polynomials of $e^{\pm i\theta}$.

By comparing the coefficients for $\dket{S_\switch}*\dket{\sigma_{\vec{r}^*_k}}*\dket{\sigma_{\vec{q}_1}}$, we find that
\begin{align}\label{eq:sum_equal_linear}
    \sum_{m\in \mathbb{F}_k}\xi_{m1}^{\{\vec{r}_l\}_l,\vec{q}_1}=\sum_{m\in \mathbb{F}_k}\xi_{m1}^{\{\vec{r}_l\}_l,\vec{q}_1}[U_m=\sigma_{\vec{r}_k^*}].
\end{align}

\vspace{2ex}

\subsection{Dependence on $V_1$}
In this part of the proof, we adopt the following shorthand notations. For any unitary operators $U, V$,
\begin{align}
	\dket{C[V]}&:=\dket{C}*\bigotimes_{i=1}^M\dket{\sigma_{\vec{r}_i}}^{I_i O_i}*\dket{V}^{I_1'O_1'}\nonumber\\
	\dket{S(U, V)}&:=\dket{ S_{\switch}}*\dket{U}*\dket{V}
	\nonumber\\
	b(\sigma_{\vec{v}},V)
	&:=
	\sum_{m\in \mathbb{F}_{\vec{v}}}
	\xi^{\{\vec{r}_l\}_l,\vec{q}_1}_{m1}[V_1=V]\, ,
\end{align}
where $\mathbb{F}_{\vec{v}}:=\{1\leq i\leq M\mid \vec{r}_i=\vec{v}\}$.

For any two $n$-qubit Pauli operators $\sigma_A, \sigma_B$, either the operator $(\sigma_A+\sigma_B)/\sqrt{2}$ or  the operator  $(\sigma_A+i\sigma_B)/\sqrt{2}$ is unitary. 
When $U:=(\sigma_A+\beta \sigma_B)/\sqrt{2}$ with $\beta\in \{1, i\}$ is unitary, the following equality holds for any $\{\vec{r}_1,\ldots,\vec{r}_M\}, \sigma_A, \sigma_B$:
\begin{align}\label{eq::sq2-AB}
	0 = 
	&\dket{C[U]}-
	\frac{1}{\sqrt{2}}
	(\dket{C[\sigma_A]}+\beta \dket{C[\sigma_B]})
	\nonumber\\
	=&\frac{1}{\sqrt{2}}
	\sum_{\vec{v}\in \{\vec{r}_1,\ldots,\vec{r}_M\}}    
	\left[
	\{b(\sigma_{\vec{v}},U)-b(\sigma_{\vec{v}},\sigma_A)\}
	\dket{S(\sigma_{\vec{v}},\sigma_A)}+
	\beta\{b(\sigma_{\vec{v}},U)-b(\sigma_{\vec{v}},\sigma_B) \}
	\dket{S(\sigma_{\vec{v}},\sigma_B)}
	\right].
\end{align}

We now calculate the inner product 
\begin{equation}
\dbraket{S(\sigma_{\vec{v}}, \sigma_A)}{S(\sigma_{\vec{v}'}, \sigma_B)} = \Tr (\sigma_B \sigma_A \sigma_{\vec{v}}  \sigma_{\vec{v}'} ) + \Tr ( \sigma_{\vec{v}} \sigma_A \sigma_B \sigma_{\vec{v}'} ) \, .
\end{equation}
From this, it is clear that $\dbraket{S(\sigma_{\vec{v}}, \sigma_A)}{S(\sigma_{\vec{v}'}, \sigma_B)} = 0$ if $\sigma_{\vec{v}'} \not \propto \sigma_{\vec{v}} \sigma_A\sigma_B$. Therefore, taking an inner product of Eq.~\eqref{eq::sq2-AB} with $\dket{S(\sigma_{\vec{v}}, \sigma_A)}$, we obtain
\begin{align}
	\begin{cases}
	  \{b(\sigma_{\vec{v}},U)-b(\sigma_{\vec{v}},\sigma_A)\} \dbraket{S(\sigma_{\vec{v}}, \sigma_A)} = 0 \textrm{~~~~if for all}~~ \sigma_{\vec{v}'} \in \{\sigma_{\vec{r}_1},\ldots,\sigma_{\vec{r}_M}\}: \sigma_{\vec{v}'} \not \propto \sigma_{\vec{v}} \sigma_A \sigma_B \, \\
	  \\
	 \{b(\sigma_{\vec{v}},U)-b(\sigma_{\vec{v}},\sigma_A)\} \dbraket{S(\sigma_{\vec{v}}, \sigma_A)} \\
	 ~~~+ \beta \{b(\gamma \sigma_{\vec{v}}\sigma_A \sigma_B,U)-b(\gamma \sigma_{\vec{v}}\sigma_A \sigma_B, \sigma_B) \} \dbraket{S(\sigma_{\vec{v}}, \sigma_A)}{S(\gamma\sigma_{\vec{v}}\sigma_A \sigma_B, \sigma_B)} = 0 \textrm{~~~~else},
	\end{cases}
\end{align}
where $\gamma\in \{1,-1, i,-i\}$ is defined by the unique choice of $\sigma_{\vec{v}'} \in \{\sigma_{\vec{r}_1},\ldots,\sigma_{\vec{r}_M}\}$  such that
	\begin{align}\label{eq:gamma}
		\sigma_{\vec{v}'} = \gamma \sigma_{\vec{v}} \sigma_A\sigma_B \, .
	\end{align}

In the first case, i.e., if for all $\sigma_{\vec{v}'} \in \{\sigma_{\vec{r}_1},\ldots,\sigma_{\vec{r}_M}\}: \sigma_{\vec{v}'} \not \propto \sigma_{\vec{v}} \sigma_A\sigma_B$, we directly obtain
\begin{align}\label{eq:UA0}
	b(\sigma_{\vec{v}},U)-b(\sigma_{\vec{v}},\sigma_A) = 0 \, .
\end{align}
In the second case, if 
\begin{equation}\label{eq:gamma_vAB}
	\dbraket{S(\sigma_{\vec{v}}, \sigma_A)}{S(\gamma \sigma_{\vec{v}}\sigma_A\sigma_B, \sigma_B)} =  \Tr (\sigma_B \sigma_A \sigma_{\vec{v}} \gamma \sigma_{\vec{v}} \sigma_A \sigma_B ) + \Tr ( \sigma_{\vec{v}} \sigma_A \sigma_B \gamma \sigma_{\vec{v}} \sigma_A \sigma_B ) = 0
\end{equation}
holds, we also obtain Eq.\ \eqref{eq:UA0}

By a similar argument, if for all $\sigma_{\vec{v}''} \in \{\sigma_{\vec{r}_1},\ldots,\sigma_{\vec{r}_M}\}: \sigma_{\vec{v}''} \not \propto \sigma_{\vec{v}} \sigma_B \sigma_A$, we directly obtain
\begin{align}\label{eq:UB0}
	b(\sigma_{\vec{v}},U)-b(\sigma_{\vec{v}},\sigma_B) = 0 \, .
\end{align}
Alternatively, if
\begin{equation}\label{eq:delta_vAB}
	\dbraket{S(\sigma_{\vec{v}}, \sigma_B)}{S(\gamma \sigma_{\vec{v}}\sigma_B\sigma_A, \sigma_A)} =  \Tr (\sigma_A \sigma_B \sigma_{\vec{v}} \delta \sigma_{\vec{v}} \sigma_B \sigma_A ) + \Tr ( \sigma_{\vec{v}} \sigma_B \sigma_A \delta \sigma_{\vec{v}} \sigma_B \sigma_A ) = 0
\end{equation}
holds, where $\delta \in \{1,-1, i,-i\}$ is defined by the unique choice of $\sigma_{\vec{v}''} \in \{\sigma_{\vec{r}_1},\ldots,\sigma_{\vec{r}_M}\}$  such that
 \begin{align}\label{eq:delta}
 	\sigma_{\vec{v}''} = \delta \sigma_{\vec{v}} \sigma_B\sigma_A \, ,
 \end{align}
we also obtain Eq.\ \eqref{eq:UB0}.

Consider now the conditions for Eq.~\eqref{eq:gamma_vAB} to be satisfied. The first term $ \Tr (\sigma_B \sigma_A \sigma_{\vec{v}} \gamma \sigma_{\vec{v}} \sigma_A \sigma_B ) = \gamma \Tr \1 $. For the second term, there are four cases:
\begin{itemize}
\item If $\gamma= \pm 1$, then $\pm \sigma_{\vec{v}} \sigma_A\sigma_B$ is an $n$-qubit Pauli operator so the second term $\Tr[\sigma_{\vec{v}} \sigma_A \sigma_B \gamma \sigma_{\vec{v}} \sigma_A \sigma_B ]=  \gamma \Tr[(\pm \sigma_{\vec{v}} \sigma_A \sigma_B) (\pm \sigma_{\vec{v}} \sigma_A \sigma_B)]  = \gamma \Tr \1 $. 
\item If $\gamma=\pm i$, then $\pm i \sigma_{\vec{v}} \sigma_A\sigma_B$ is an $n$-qubit Pauli operator so the second term $\Tr[(\sigma_{\vec{v}} \sigma_A \sigma_B) \gamma (\sigma_{\vec{v}} \sigma_A \sigma_B)]  = - \gamma \Tr[(\pm i \sigma_{\vec{v}} \sigma_A \sigma_B)  (\pm i \sigma_{\vec{v}} \sigma_A \sigma_B)] = - \gamma \Tr \1 $. 
\end{itemize}
Therefore, Eq.~\eqref{eq:gamma_vAB} is satisfied if and only if $\gamma = \pm i$. By a similar argument,  Eq.\ \eqref{eq:delta_vAB} is satisfied if and only if $\delta = \pm i$. 

Note that the following equivalences hold: $\forall \sigma_{\vec{v}''} \in \{\sigma_{\vec{r}_1},\ldots,\sigma_{\vec{r}_M}\}: \sigma_{\vec{v}''} \not \propto \sigma_{\vec{v}} \sigma_B \sigma_A \iff \forall \sigma_{\vec{v}'} \in \{\sigma_{\vec{r}_1},\ldots,\sigma_{\vec{r}_M}\}: \sigma_{\vec{v}'} \not \propto \sigma_{\vec{v}} \sigma_A\sigma_B$, and also $\gamma \in \{i, -i\} \iff \delta \in \{i, -i\} $. Therefore,  for all $ \{\sigma_{\vec{r}_1},\ldots,\sigma_{\vec{r}_M}\}$, for every tuple $(  \sigma_{\vec{v}} \in\{\sigma_{\vec{r}_1},\ldots,\sigma_{\vec{r}_M}\},  \sigma_A, \sigma_B)$, if one of the two following conditions is satisfied:
\begin{enumerate}
\item  $ \forall \, \sigma_{\vec{v}'} \in \{\sigma_{\vec{r}_1},\ldots,\sigma_{\vec{r}_M}\}: \sigma_{\vec{v}'} \not \propto \sigma_{\vec{v}} \sigma_A\sigma_B$, or
\item  $\exists \, \sigma_{\vec{v}'}$ such that $ \sigma_{\vec{v}'} = \pm i \sigma_{\vec{v}} \sigma_A \sigma_B$, 
\end{enumerate}
then 
\begin{align}\label{eq:AUB_equal}
	b(\sigma_{\vec{v}}, \sigma_A) = b(\sigma_{\vec{v}}, U) \, ,\\
  b(\sigma_{\vec{v}}, \sigma_B) = b(\sigma_{\vec{v}}, U) \, ,
\end{align}
which implies that 
\begin{align}\label{eq:AB_equal}
	b(\sigma_{\vec{v}}, \sigma_A) = b(\sigma_{\vec{v}}, \sigma_B)  \, .
\end{align}
We now consider the choices of $(\sigma_{\vec{v}}, \sigma_A,  \sigma_B)$ where neither of the above two conditions is satisfied.

\textit{The case where  $\sigma_{\vec{v}} = \sigma_A$ or $\sigma_{\vec{v}} = \sigma_B$:} First, note that if $\sigma_{\vec{v}} = \sigma_B$, and Condition 1. is not satisfied, then Eq.\ \eqref{eq:gamma}  implies that $\sigma_{\vec{v}'} = \sigma_A $ and $\gamma = \pm 1$, so Condition 2. is not satisfied either. In this case, we can take another $n$-qubit Pauli operator $\sigma_C \notin \{\sigma_{\vec{r}_1},\ldots,\sigma_{\vec{r}_M}\}$ (which implies that $\sigma_C \neq \sigma_A, \sigma_B$), such that $\sigma_B  \sigma_A \sigma_C = \pm i \sigma_{BAC}$ for some $n$-qubit Pauli $\sigma_{BAC}$. The existence of such a $ \sigma_C$ is guaranteed by:
\begin{itemize}
\item the fact that half of the total $4^n$ number of $n$-qubit Pauli operators, when multiplied after an $n$-qubit Pauli operator (in this case $(\gamma' \sigma_B \sigma_A)$ with $\gamma' \in \{1,-1, i, -i\}$), gives a Pauli operator times $\pm 1$ and the other half will give a Pauli operator times $\pm i$, 
\item the fact that the set $\{\sigma_{\vec{r}_1},\ldots,\sigma_{\vec{r}_M}\}$ contains the operators $\sigma_A, \sigma_B$, which, when multiplied after the $n$-qubit Pauli operator $(\gamma' \sigma_B \sigma_A)$, gives a Pauli operator times $\pm 1$,
\item  the assumption that $M < 4^n/2 + 2$.
\end{itemize}
Applying the procedure in Eqs.\ \eqref{eq::sq2-AB}--\eqref{eq:AB_equal} above to the unitary $U' := (\sigma_C + \beta' \sigma_B)/\sqrt{2}$ (with $\beta' \in 
\{1,i\}$), we find that there is no $\sigma_{\vec{v}'} \in \{\sigma_{\vec{r}_1},\ldots,\sigma_{\vec{r}_M}\}$ such that $\sigma_{\vec{v}'} \propto \sigma_{\vec{v}} \sigma_C \sigma_B = \sigma_{B} \sigma_C \sigma_B \propto  \sigma_C  $. 
Therefore, Condition 1. is satisfied for $(  \sigma_{\vec{v}} \in\{\sigma_{\vec{r}_1},\ldots,\sigma_{\vec{r}_M}\},  \sigma_C, \sigma_B)$ and we have that
\begin{equation}
	b(\sigma_{\vec{v}}, \sigma_C) = b(\sigma_{\vec{v}}, U') = b(\sigma_{\vec{v}}, \sigma_B) \, .	
\end{equation}
Applying the procedure in Eqs.\ \eqref{eq::sq2-AB}--\eqref{eq:AB_equal} above to the unitary $U'' := (\sigma_A + \beta'' \sigma_C)/\sqrt{2}$ (with $\beta'' \in 
\{1,i\}$), we find that either (a) there is no $\sigma_{\vec{v}''} \in \{\sigma_{\vec{r}_1},\ldots,\sigma_{\vec{r}_M}\}$ such that $\sigma_{\vec{v}''} \propto \sigma_{\vec{v}} \sigma_A \sigma_C$, or (b) if there is, then $ \sigma_{\vec{v}} \sigma_A \sigma_C = \sigma_{B} \sigma_A \sigma_C = \pm i \sigma_{BAC}$, i.e. $\sigma_{\vec{v}''} = \sigma_{BAC}$. Therefore, for $(  \sigma_{\vec{v}} \in\{\sigma_{\vec{r}_1},\ldots,\sigma_{\vec{r}_M}\},  \sigma_A, \sigma_C)$,  either Condition 1. or 2. is satisfied and we have that
\begin{equation}
	b(\sigma_{\vec{v}}, \sigma_A) = b(\sigma_{\vec{v}}, U'') = b(\sigma_{\vec{v}}, \sigma_C) \, .	
\end{equation}
Overall, Eq.\ \eqref{eq:AB_equal} is satisfied for $(  \sigma_{\vec{v}} \in\{\sigma_{\vec{r}_1},\ldots,\sigma_{\vec{r}_M}\},  \sigma_A, \sigma_B)$ .

\textit{The case where  $\sigma_{\vec{v}} \neq \sigma_A, \sigma_B$:}
If neither Condition 1.\ nor Condition 2.\ are satisfied, then we can take another $n$-qubit Pauli operator 
$\sigma_C \neq \sigma_A, \sigma_B$,
such that 
\begin{align}
\sigma_{\vec{v}}  \sigma_A \sigma_C &= \pm i \sigma_{vAC} \, , \\
\sigma_{\vec{v}}  \sigma_C \sigma_B &\in \{ \pm i \sigma_{vCB} \} \iff \sigma_{\vec{v}}  \sigma_B \sigma_C \in \{ \pm i \sigma_{vCB} \} \,	,
\end{align}
for some $n$-qubit Pauli operators $\sigma_{vAC},\sigma_{vCB}$. 
The existence of such a $ \sigma_C$ is guaranteed by:
\begin{itemize}	
	\item the fact that for any two different non-identity $n$-qubit Pauli operators $\sigma_{E},\sigma_{F}$, there exists an $n$-qubit Pauli operator $\sigma_{ Q}$ such that both $\sigma_{E}\sigma_{  Q}$ and $\sigma_{F}\sigma_{\rm  Q}$ are equal to $+i$ or $-i$ times an $n$-qubit Pauli operator, 
	\item the fact that for any two different non-identity  $n$-qubit Pauli operators $\sigma_{E},\sigma_{F}$, there exists an $n$-qubit Pauli operator $\sigma_{ R}$ such that $\sigma_{E}\sigma_{  R}$ equals $\pm i$  times an $n$-qubit Pauli operator, while $\sigma_{F}\sigma_{  R}$ equals $\pm 1$ times an $n$-qubit Pauli operator.
\end{itemize}
This enables $\sigma_C$ to be chosen according to the following strategy:
\begin{itemize}
	\item if there are $\sigma_{E}, \sigma_{F}$ such that  $\sigma_{E} = \pm  \sigma_{v} \sigma_{A}$ and $\sigma_{F}= \pm \sigma_{v} \sigma_{B}$, then pick $\sigma_C = \sigma_{  Q}$ as defined above,
	\item if there are $\sigma_{E}, \sigma_{F}$ such that either $\sigma_{E} = \pm  \sigma_{v} \sigma_{A}$ and $\sigma_{F}= \pm i  \sigma_{v} \sigma_{B}$, or $\sigma_{F} = \pm i \sigma_{v} \sigma_{A}$ and $\sigma_{E}= \pm   \sigma_{v} \sigma_{B}$, then pick $\sigma_C = \sigma_{  R}$ as defined above,
	\item if there are $\sigma_{E}, \sigma_{F}$ such that $\sigma_{E} = \pm i \sigma_{v} \sigma_{A}$ and $\sigma_{F}= \pm i \sigma_{v} \sigma_{B}$, then pick $\sigma_C =  \sigma_{F}$, in which case $\sigma_{\vec{v}}  \sigma_B \sigma_C =  \sigma_{\vec{v}} \sigma_B (\pm i \sigma_{v} \sigma_{B}) = \mp i \1$ and $\sigma_{\vec{v}}  \sigma_A \sigma_C = \pm i \sigma_{\vec{v}}  \sigma_A \sigma_{\vec{v}} \sigma_B = \mp i \sigma_{\vec{v}} \sigma_{\vec{v}} \sigma_A  \sigma_B =  \mp i \sigma_{\vec{v}} \sigma_B \sigma_A  \sigma_{\vec{v}} = - \sigma_C  \sigma_A \sigma_{\vec{v}} = - (\sigma_{\vec{v}}  \sigma_A \sigma_C)^\dag$ (where in the third equality we use the assumption that Conditions 1.\ and 2.\ are not satisfied, which implies that $\sigma_{\vec{v}} \sigma_A  \sigma_B = \sigma_B \sigma_A  \sigma_{\vec{v}}$), and therefore $\sigma_{\vec{v}}  \sigma_A \sigma_C$ must be proportional to $\pm i$ times a Pauli.
\end{itemize}
Applying the procedure in Eqs.\ \eqref{eq::sq2-AB}--\eqref{eq:AB_equal} above to the unitary $U' := (\sigma_C + \beta' \sigma_B)/\sqrt{2}$ (with $\beta' \in 
\{1,i\}$), we find that Condition 1.\ or 2.\ is satisfied for $(  \sigma_{\vec{v}} \in\{\sigma_{\vec{r}_1},\ldots,\sigma_{\vec{r}_M}\},  \sigma_C, \sigma_B)$ and we have that
\begin{equation}
	b(\sigma_{\vec{v}}, \sigma_C) = b(\sigma_{\vec{v}}, U') = b(\sigma_{\vec{v}}, \sigma_B) \, .	
\end{equation}
Applying the procedure in Eqs.\ \eqref{eq::sq2-AB}--\eqref{eq:AB_equal} above to the unitary $U'' := (\sigma_A + \beta'' \sigma_C)/\sqrt{2}$ (with $\beta'' \in 
\{1,i\}$), we find that Condition 1.\ or 2.\ is satisfied for $(  \sigma_{\vec{v}} \in\{\sigma_{\vec{r}_1},\ldots,\sigma_{\vec{r}_M}\},  \sigma_A, \sigma_C)$ and we have that
\begin{equation}
	b(\sigma_{\vec{v}}, \sigma_A) = b(\sigma_{\vec{v}}, U'') = b(\sigma_{\vec{v}}, \sigma_C) \, .	
\end{equation}
Overall, Eq.\ \eqref{eq:AB_equal} is satisfied for  $(  \sigma_{\vec{v}} \in\{\sigma_{\vec{r}_1},\ldots,\sigma_{\vec{r}_M}\},  \sigma_A, \sigma_B)$.

Having shown that for all $ \{\sigma_{\vec{r}_1},\ldots,\sigma_{\vec{r}_M}\}$, for every tuple $(  \sigma_{\vec{v}} \in\{\sigma_{\vec{r}_1},\ldots,\sigma_{\vec{r}_M}\},  \sigma_A, \sigma_B)$, Eq.\ \eqref{eq:AB_equal} is satisfied, we conclude that $b(\sigma_{\vec{v}},\sigma_{\vec{q_1}}) :=
	\sum_{m\in \mathbb{F}_{\vec{v}}}
	\xi^{\{\vec{r}_l\}_l,\vec{q}_1}_{m1}$ is independent of the choice of $\sigma_{\vec{q_1}}$. This means that 
 \begin{equation}\label{eq:V1_indep}
    \sum_{m\in \mathbb{F}_{\vec{v}}}
	\xi^{\{\vec{r}_l\}_l,\vec{q}_1}_{m1} = \sum_{m\in \mathbb{F}_{\vec{v}}}
	\xi^{\{\vec{r}_l\}_l,\vec{q}_1}_{m1} [V_1= \sigma_{\vec{q}^*}] \, ,
 \end{equation}
for any $n$-qubit Pauli operator $\sigma_{\vec{q}^*}$.

\subsection{Proving that the redefinition of Eq.~\eqref{eq:global_def_M1} satisfies Eq.\ \eqref{eq:indep_M1}}

Equations\ \eqref{eq:sum_equal_linear} and Eq.\ \eqref{eq:V1_indep} together show that for all $\vec{r}_1,\ldots,\vec{r}_M, \vec{q_1} \in \{0,1,2,3\}^{\times n}$,
\begin{align}
    \dket{C}*\bigotimes_{i=1}^M \dket{\sigma_{\vec{r}_i}}^{I_iO_i}*\dket{\sigma_{\vec{q}_1}}^{I_1'O_1'} 
    &=  \sum_{k=1}^M \xi^{\{\vec{r}_l\}_l,\vec{q}_1}_{k1} \dket{ S_{\switch}}*\dket{\sigma_{\vec{r}_k}}*\dket{\sigma_{\vec{q}_1}} 
\nonumber\\
   &=\sum_{\vec{v}\in \{\vec{r}_1,\ldots,\vec{r}_M\}}
    \sum_{m\in \mathbb{F}_{\vec{v}}}
    \xi^{\{\vec{r}_l\}_l,\vec{q}_1}_{m1}
    \dket{ S_{\switch}}*\dket{\sigma_{\vec{v}}}*\dket{\sigma_{\vec{q}_1}}
\nonumber\\
   &=
   \sum_{\vec{v}\in \{\vec{r}_1,\ldots,\vec{r}_M\}}
    \sum_{m\in \mathbb{F}_{\vec{v}}}
    \xi^{\{\vec{r}_l\}_l,\vec{q}_1}_{m1} [V_1= \sigma_{\vec{q}^*}]
    \dket{ S_{\switch}}*\dket{\sigma_{\vec{v}}}*\dket{\sigma_{\vec{q}_1}}
\nonumber\\
    &=
    \sum_{\vec{v}\in \{\vec{r}_1,\ldots,\vec{r}_M\}}
    \sum_{m\in \mathbb{F}_{\vec{v}}}\xi^{\{\vec{r}_l\}_l,\vec{q}_1}_{m1}[U_m=\sigma_{\vec{v}^*}, V_1= \sigma_{\vec{q}^*}]
    \dket{ S_{\switch}}*\dket{\sigma_{\vec{v}}}*\dket{\sigma_{\vec{q}_1}}
\nonumber\\
    &=\sum_{k=1}^M\tilde{\xi}_{k1}(\{U_i=\sigma_{\vec{r}_i}\}_{i=1}^M,V_1=\sigma_{\vec{q}_1})\dket{ S_{\switch}}*\dket{\sigma_{\vec{r}
_k}}*\dket{\sigma_{\vec{q}_1}}
   \, ,
\end{align}
where $\mathbb{F}_{\vec{v}}:=\{1\leq i\leq M\mid \vec{r}_i=\vec{v}\}$, $\vec{v}^* \in \{0,1,2,3\}^{\times n}$ is an arbitrary vector outside of the set $\{\vec{r}_1,\ldots,\vec{r}_M\} \backslash \{\vec{v}\}$, and $\vec{q}^* \in \{0,1,2,3\}^{\times n}$ is an arbitrary fixed vector.
Therefore, $\tilde{\xi}_{k1}$ as defined in Eq.~(\ref{eq:global_def_M1}) indeed satisfies Eq.~(\ref{eq:indep_M1}).

This also implies that 
\begin{align}
    &\dket{C} *  \bigotimes_{i=1}^M
\dket{\sum_{\vec{r}_i}\alpha^i_{\vec{r}_i}\sigma_{\vec{r}_i}}^{I_{i} O_{i}} 
 * \dket{\sum_{\vec{q}_1}\beta^1_{\vec{q}_1}\sigma_{\vec{q}_1}
 }^{I'_1 O'_1} 
 =
 \sum_{\{\vec{r}_l\}_l, \vec{q}_1}\left(\prod_{j=1}^M \alpha^j_{\vec{r}_j}\right)\beta_{\vec{q}_1}^1 \dket{C}* \bigotimes_{i=1}^M
  \dket{\sigma_{\vec{r}_{i}}}^{I_{i} O_{i}} 
 * \dket{\sigma_{\vec{q}_1}}^{I'_1 O'_1} 
 \nonumber\\
 &=
 \sum_{\{\vec{r}_l\}_l, \vec{q}_1}
 \left(\prod_{j=1}^M \alpha^j_{\vec{r}_j}\right)\beta_{\vec{q}_1}^1
 \left[
 \sum_{k=1}^M \tilde{\xi}_{k1}(\{U_m=\sigma_{\vec{r}_m}\}_m,V_1=\sigma_{\vec{q}_1}) \dket{S_\switch} * \dket{\sigma_{\vec{r}_k}}^{I_k O_k} * \dket{\sigma_{\vec{q}_1}}^{I'_1 O'_1}
 \right]
\nonumber\\
&=
\sum_{k=1}^M
\left[
\sum_{\{\vec{r}_l\}_{l\neq k}}\left(
\prod_{j=1| j \neq k}^{M}\alpha_{\vec{r}_j}^j
\right) \tilde{\xi}_{k1}(\{U_m=\sigma_{\vec{r}_m}\}_m,V_1=\sigma_{\vec{q}_1}) \right]
\sum_{\vec{r}_k, \vec{q}_1}
\alpha^{k}_{\vec{r}_k}
\beta^1_{\vec{q}_1}
\dket{ S_{\switch}}*
\dket{
\sigma_{\vec{r}_k}
}^{I_kO_k}*
\dket{\sigma_{\vec{q}_1}}^{I_1'O_1'}
\nonumber\\
 &=\sum_{k=1}^M \tilde{\xi}_{k1}\left(\left\{U_i=\sum_{\vec{r}_i}{\alpha_{\vec{r}_i}^i \sigma_{\vec{r}_i}}\right\},\ V_1=\sum_{\vec{q}_1}\beta_{\vec{q}_1}^1\sigma_{\vec{q}_1}\right) \dket{S_\switch} * \dket{\sum_{\vec{r}_k}\alpha_{\vec{r}_k}^k\sigma_{\vec{r}_k}}^{I_k O_k} * \dket{\sum_{\vec{q}_1}\beta^1_{\vec{q}_1}\sigma_{\vec{q}_1}}^{I'_1 O'_1} \, ,
\end{align}
i.e.,\ Eq.\ \eqref{eq:global_def_M1_linear} also satisfies Eq.~\eqref{eq:indep_M1}.
\end{proof}

\vspace{2ex}

\vspace{3ex}

%%%%%%%%%%%%%%%%%%%%%%%%%%%%%%%%%%%%%

\begin{lemma}\label{lem:p_as_vector_MN}
Let $C \in \mbL(I_1 \otimes \cdots \otimes I_M \otimes O_1 \otimes \cdots \otimes O_M \otimes I'_1 \otimes \cdots \otimes I'_N \otimes O'_1 \otimes \cdots \otimes O'_N \otimes P_C \otimes P_T \otimes F_C \otimes F_T) $, for some $M, N \in \mathbb{N}^+$ where $P_T, F_T, \{I_i\}_{i}, \{O_i\}_{i}, \{I'_j\}_{j}, \{O'_j\}_{j}$ correspond to  $n$-qubit Hilbert spaces for some $n \in \mathbb{N}^+$, and $P_C, F_C$ correspond to  qubit Hilbert spaces, be a linear operator such that $C = \dketbra{C}{C}$ for some vector $\dket{C}$. 
If, for a given $M,N$, 
for all $(M+N)$-tuples of $n$-qubit unitary operators $(U_1, \dots, U_M, V_1, \dots, V_N)$,
\begin{equation}\label{eq:indep}
    \dket{C} * \dket{U_1}^{I_1 O_1} * \dots * \dket{U_M}^{I_M O_M} *  \dket{V_1}^{I'_1 O'_1} * \dots * \dket{V_N}^{I'_N O'_N} = \sum_{k=1}^M \sum_{l=1}^N  \tilde{\xi}_{kl} \dket{S_\switch} * \dket{U_k}^{I_k O_k} * \dket{V_l}^{I'_l O'_l}
\end{equation}
for some complex numbers $\tilde{\xi}_{kl}:=\tilde{\xi}_{kl}(\{U_i\}_i,\ \{V_j\}_j) \in \mathbb{C}$, for all $k \in \{1, \dots, M\}$, $l \in \{1, \dots, N\}$,  
which are simultaneously
\begin{enumerate}
    \item  independent of $U_{k}$ and $V_l$, and 
    \item  linear in $U_{i}$ and $V_{j}$ for all $i \neq k, j \neq l$,
\end{enumerate}
then 
\begin{align} \label{eq:pure_C_lemmapasvector}
      &\dket{C}^{PF, I_1 O_1, I'_1 O'_1, \cdots , I_M O_M , I'_N O'_N }
= \sum_{k=1}^M \sum_{l=1}^N 
\dket{S_\switch}^{PF I_k O_k I'_l O'_l} \otimes
        \dket{\tilde{\xi}_{kl}}^{\{I_1 O_1, I'_1 O'_1, \cdots , I_M O_M , I'_N O'_N\} \setminus \{I_k O_k, I'_l O'_l\} } 
\end{align}
for some vectors $\dket{\tilde{\xi}_{kl}}^{\{I_1 O_1, I'_1 O'_1, \cdots , I_M O_M , I'_N O'_N\} \setminus \{I_k O_k, I'_l O'_l\} }$ (that are independent of $\{U_i\}_i$ and $\{V_j\}_j$).
\end{lemma}

\begin{proof}
The first (independence) condition implies that we can write $\tilde{\xi}_{kl}(\{U_i\}_i,\ \{V_j\}_j) = \tilde{\xi}_{kl}(\{U_i\}_{i \neq k}^M,\ \{V_j\}_{j \neq l}^N)$.
The second (linearity) condition implies that we can write the linear functions $\tilde{\xi}_{kl}(\{U_i\}_{i \neq k}^M,\ \{V_j\}_{j \neq l}^N)$ using vectors $\dket{\tilde{\xi}_{kl}}$ by
\begin{align}
    \tilde{\xi}_{kl}(\{U_i\}_{i \neq k}^M,\ \{V_j\}_{j \neq l}^N)
    \eqqcolon 
    \dket{\tilde{\xi}_{kl}}\ast \bigotimes_{i \neq k}^M \dket{U_i} \otimes
        \bigotimes_{j \neq l}^N \dket{V_j} \, .
\end{align}
Then, by explicitly writing in the system labels, Eq.~(\ref{eq:indep}) becomes:
For any sets of $n$-qubit unitaries $\{U_i\}_{i=1}^M$ and $\{V_j\}_{j=1}^N$, 
\begin{align}
      &\dket{C}^{PF, I_1 O_1, I'_1 O'_1, \cdots , I_M O_M , I'_N O'_N } 
      \ast   
         \left[
         \bigotimes_{i=1}^M \dket{U_i}^{I_i O_i} \otimes
          \bigotimes_{j=1}^N \dket{V_j}^{I'_j O'_j}
          \right]
\\          
&= \left[ \sum_{k= 1}^M \sum_{l= 1}^N 
        \dket{\tilde{\xi}_{kl}}^{\{I_1 O_1, I'_1 O'_1, \cdots , I_M O_M , I'_N O'_N\} \setminus \{I_k O_k, I'_l O'_l\}}  \otimes 
          \dket{S_\switch}^{PF I_k O_k I'_l O'_l}
          \right]
          \ast 
          \left[
       \bigotimes_{i=1}^M  \dket{U_i}^{I_i O_i} \otimes
       \bigotimes_{j=1}^N  \dket{V_j}^{I'_j O'_j}
        \right] \, , \nonumber
\end{align}
for some vectors $\dket{\tilde{\xi}_{kl}}^{\{I_1 O_1, I'_1 O'_1, \cdots , I_M O_M , I'_N O'_N\} \setminus \{I_k O_k, I'_l O'_l\} }$.

Since the equation is true for all unitaries $\{U_i\}$ and $\{V_j\}$,  and  $\mathrm{span}(\{\dket{U}\,|\,U\in\text{SU}(d)\})=\mathbb{C}^d\otimes\mathbb{C}^d$, it implies that:
    \begin{align}
      &\dket{C}^{PF, I_1 O_1, I'_1 O'_1, \cdots , I_M O_M , I'_N O'_N }
=  \sum_{k= 1}^M \sum_{l= 1}^N 
\dket{S_\switch}^{PF I_k O_k I'_l O'_l} \otimes
        \dket{\tilde{\xi}_{kl}}^{\{I_1 O_1, I'_1 O'_1, \cdots , I_M O_M , I'_N O'_N\} \setminus \{I_k O_k, I'_l O'_l\} } 
    \end{align}
    for some vectors $\dket{\tilde{\xi}_{kl}}^{\{I_1 O_1, I'_1 O'_1, \cdots , I_M O_M , I'_N O'_N\} \setminus \{I_k O_k, I'_l O'_l\} }$.
\end{proof}

\vspace{3ex}

\begin{lemma}\label{lem:qccc_invalid}
    Suppose $C$ is the Choi matrix of an $(M+N)$-slot QC-CC supermap \cite{wechs2021quantum}, i.e., it satisfies the QC-CC conditions given by:
    \begin{align}
    	&C= \sum_{\vec{r}_{M+N} \in \mathrm{Perm}(1, \ldots, M+N)} C_{P \vec{r}_{M+N} F},\\
    	&\textup{such that}\quad\quad
        C_{P \vec{r}_{M+N} F} \geq 0 \quad \forall \vec{r}_{M+N},\\
        &\Tr_{F}[C_{P \vec{r}_{M+N} F}] = C_{P \vec{r}_{M+N}} \otimes \1^{O_{r_{M+N}}} \quad \forall \vec{r}_{M+N},\label{eq:qc-cc_F}\\
        &\sum_{r_{m+1}}\Tr_{I_{r_{m+1}}}[C_{P \vec{r}_m r_{m+1}}] = C_{P \vec{r}_m} \otimes \1^{O_{r_{m}}} \quad \forall m\in\{1, \ldots, M+N-1\}, \forall \vec{r}_m \coloneqq (r_1, \ldots, r_m),\label{eq:qc-cc_m}\\
        &\sum_{r_1} \mathrm{tr}_{I_{r_1}}[C_{P r_1}] = \1^{P},\label{eq:qc-cc_norm}
    \end{align}
    where the dimension of input and output spaces are $d$, i.e., $\mathcal{H}^{I_i} \cong \mathcal{H}^{O_j} \cong \mathbb{C}^d$ for all $i, j\in\{1, \ldots, M+N\}$, and $\vec{r}_m r_{m+1}$ represents a vector $(r_1, \ldots, r_m, r_{m+1})$, with
    each vector $\vec{r}_m$ composed of elements $r_1, \ldots , r_m$.
    We rename the last $N$ input and output systems as
    \begin{align}
        I'_k \coloneqq I_{M + k}, \quad O'_k \coloneqq O_{M+k} \quad \forall k\in\{1, \ldots, N\}. 
    \end{align}
    The operators $C_{P\vec{r}_{m}} \in \mathbb{L}(I_1\otimes \cdots \otimes I_{m} \otimes O_1\otimes \cdots \otimes O_{m-1}\otimes P_C\otimes P_T)$ for $m\in\{1, \ldots, M+N\}$ are recursively defined by
    \begin{align}
        C_{P\vec{r}_{M+N}} &\coloneqq {1\over d} \Tr_{O_{r_{M+N}} F}[C_{P\vec{r}_{M+N} F}],\\
        C_{P\vec{r}_m}&\coloneqq {1\over d} \sum_{r_{m+1}}\Tr_{O_{r_m} I_{r_{m+1}}}[C_{P\vec{r}_m r_{m+1}}] \quad \forall m\in\{1, \ldots, M+N-1\}.
    \end{align}
    If $\max(M, N)\leq  \max(2, d-1)$ holds, then the set of Choi matrices $\{C_{P\vec{r}_{M+N} F}\}_{\vec{r}_{M+N}}$ cannot be in the form given by
    \begin{align}
        C_{P\vec{r}_{M+N} F} &= \sum_{a}\dketbra{C_{P\vec{r}_{M+N} F}^{(a)}},\\
        \dket{C_{P\vec{r}_{M+N} F}^{(a)}} &= \sum_{i=1}^{M}\sum_{k=1}^{N} \dket{S_\switch}^{I_i O_i I'_k O'_k P_C P_T F_C F_T} \otimes \dket{\tilde{\xi}_{ik}^{(a),\vec{r}_{M+N}}},
    \end{align}
    where $\dketbra{S_\switch}$ is the Choi matrix of the quantum switch and $\dket{\tilde{\xi}_{ik}^{(a),\vec{r}_{M+N}}} \in \mathcal{H}^{I_{\bar{i}}} \otimes \mathcal{H}^{O_{\bar{i}}} \otimes \mathcal{H}^{I'_{\bar{k}}} \otimes \mathcal{H}^{O'_{\bar{k}}}$ for $i\in\{1, \ldots, M\}$ and $k\in \{1, \ldots, N\}$, where $\mathcal{H}^{I_{\bar{i}}} \coloneqq \bigotimes_{i'\neq i} \mathcal{H}^{I_{i'}}, \mathcal{H}^{O_{\bar{i}}} \coloneqq \bigotimes_{i'\neq i} \mathcal{H}^{O_{i'}}, \mathcal{H}^{I'_{\bar{k}}} \coloneqq \bigotimes_{k'\neq k} \mathcal{H}^{I'_{k'}}, \mathcal{H}^{O'_{\bar{k}}} \coloneqq \bigotimes_{k'\neq k} \mathcal{H}^{O'_{k'}}$.
\end{lemma}

\begin{proof}
    We assume that the set $\{C_{P\vec{r}_{M+N} F}\}_{\vec{r}_{M+N}}$ forms a QC-CC supermap and show a contradiction to complete the proof.
    To this end, we use Eqs.~\eqref{eq:qc-cc_F} and \eqref{eq:qc-cc_m} in the QC-CC conditions to show the following equation for $C_{P \vec{r}_m}$:
    \begin{align}
    \label{eq:C_Prm}
        \sum_{r_m}\Tr_{I_{r_{m}}} [C_{P \vec{r}_m}]
        &=\sum_{i,j\in \mbA_{\vec{r}_{m-1}}} \sum_{k,l\in \mbB_{\vec{r}_{m-1}}} (\ketbra{0}^{P_C} \otimes \dket{\1}^{P_T I_i}\dbra{\1}^{P_T I_j} \otimes \dket{\1}^{O_i I'_k}\dbra{\1}^{O_j I'_l} \otimes \1^{O'_l \to O'_k} \nonumber\\
        &+ \ketbra{1}^{P_C} \otimes \dket{\1}^{P_T I'_k}\dbra{\1}^{P_T I'_l} \otimes \dket{\1}^{O'_k I_i}\dbra{\1}^{O'_l I_j} \otimes \1^{O_j \to O_i}) \otimes C_{P\vec{r}_{m-1}}^{(ijkl)} \quad \forall m\in\{1, \ldots, M+N+1\},
    \end{align}
    where the summation over $r_m$ for $m=M+N+1$ is taken as $I_{r_{M+N+1}} \coloneqq F$, $C_{P\vec{r}_{M+N+1}}$ is defined by $C_{P\vec{r}_{M+N+1}}\coloneqq C_{P\vec{r}_{M+N}F}$, the set of indices $\mbA_{\vec{r}_{m-1}}$ and $\mbB_{\vec{r}_{m-1}}$ are defined by
    \begin{align}
        \mbA_{\vec{r}_{m-1}}&\coloneqq \{r_1, \ldots, r_{m-1}\} \cap \{1, \ldots, M\},\\
        \mbB_{\vec{r}_{m-1}}&\coloneqq \{r_1-M, \ldots, r_{m-1}-M\} \cap \{1, \ldots, N\},
    \end{align}
    and $C_{P\vec{r}_{m-1}}^{(ijkl)}$ is an operator.
    If this equation holds, since $\mbA_{\vec{r}_0}$ and $\mbB_{\vec{r}_0}$ are the empty sets, we obtain 
    \begin{align}
        \sum_{r_1} \Tr_{I_{r_1}}[C_{P r_1}] = 0,
    \end{align}
    which contradicts with the normalization condition \eqref{eq:qc-cc_norm} in the QC-CC conditions.
    In the rest of the proof, we show Eq.~\eqref{eq:C_Prm} by induction with respect to $m$.

    First, we show Eq.~\eqref{eq:C_Prm} for $m=M+N+1$ as follows.
    Since the operator $C_{P\vec{r}_{M+N} F}$ can be written as
    \begin{align}
        C_{P\vec{r}_{M+N} F} &= \sum_{i,j,k,l} \dket{S_\switch}^{I_i O_i I'_k O'_k P_C P_T F_C F_T}\dbra{S_\switch}^{I_j O_j I'_l O'_l P_C P_T F_C F_T} \otimes C_{P\vec{r}_{M+N}}^{(ijkl)},
    \end{align}
    where $C_{P\vec{r}_{M+N}}^{(ijkl)} \coloneqq \sum_{a} \dketbra{\tilde{\xi}_{ik}^{(a),\vec{r}_{M+N}}}{\tilde{\xi}_{jl}^{(a),\vec{r}_{M+N}}}$.
    The partial trace $\Tr_{F} C_{P\vec{r}_{M+N} F}$ is given by
    \begin{align}
        \Tr_{F} [C_{P\vec{r}_{M+N} F}] &= \sum_{i,j=1}^{M} \sum_{k,l=1}^{N} (\ketbra{0}^{P_C} \otimes \dket{\1}^{P_T I_i}\dbra{\1}^{P_T I_j} \otimes \dket{\1}^{O_i I'_k}\dbra{\1}^{O_j I'_l} \otimes \1^{O'_l \to O'_k} \nonumber\\
        &+ \ketbra{1}^{P_C} \otimes \dket{\1}^{P_T I'_k}\dbra{\1}^{P_T I'_l} \otimes \dket{\1}^{O'_k I_i}\dbra{\1}^{O'_l I_j} \otimes \1^{O_j \to O_i}) \otimes C_{P\vec{r}_{M+N}}^{(ijkl)},\label{eq:C_r}
    \end{align}
    i.e., Eq.~\eqref{eq:C_Prm} holds for $m=M+N+1$.

    To complete the proof, we show Eq.~\eqref{eq:C_Prm} by assuming Eq.~\eqref{eq:C_Prm} for $m \gets m+1$, i.e.,
    \begin{align}
    \label{eq:induction_m+1}
    \sum_{r_{m+1}}\Tr_{I_{r_{m+1}}} [C_{P \vec{r}_{m} r_{m+1}}]
        &=\sum_{i,j\in \mbA_{\vec{r}_{m}}} \sum_{k,l\in \mbB_{\vec{r}_{m}}} (\ketbra{0}^{P_C} \otimes \dket{\1}^{P_T I_i}\dbra{\1}^{P_T I_j} \otimes \dket{\1}^{O_i I'_k}\dbra{\1}^{O_j I'_l} \otimes \1^{O'_l \to O'_k} \nonumber\\
        &+ \ketbra{1}^{P_C} \otimes \dket{\1}^{P_T I'_k}\dbra{\1}^{P_T I'_l} \otimes \dket{\1}^{O'_k I_i}\dbra{\1}^{O'_l I_j} \otimes \1^{O_j \to O_i}) \otimes C_{P\vec{r}_{m}}^{(ijkl)}.
    \end{align}
    
    By symmetry with $(I_i, O_i)$ and $(I'_k, O'_k)$, it is sufficient to show if $r_{m} \in \{1, \ldots, M\}$ holds. 
    From Eq.~\eqref{eq:qc-cc_m} [or Eq.~\eqref{eq:qc-cc_F} for $m=M+N$] in the QC-CC conditions and Eq.~\eqref{eq:induction_m+1}, we obtain
    \begin{align}
    \label{eq:apply_qccc_condition_1}
        &\sum_{i,j\in \mbA_{\vec{r}_{m}}} \sum_{k,l\in \mbB_{\vec{r}_{m}}} (\ketbra{0}^{P_C} \otimes \dket{\1}^{P_T I_i}\dbra{\1}^{P_T I_j} \otimes \dket{\1}^{O_i I'_k}\dbra{\1}^{O_j I'_l} \otimes \1^{O'_l \to O'_k} \nonumber\\
        &+ \ketbra{1}^{P_C} \otimes \dket{\1}^{P_T I'_k}\dbra{\1}^{P_T I'_l} \otimes \dket{\1}^{O'_k I_i}\dbra{\1}^{O'_l I_j} \otimes \1^{O_j \to O_i}) \otimes C_{P\vec{r}_{m}}^{(ijkl)}\nonumber\\
        &=\sum_{i,j\in \mbA_{\vec{r}_{m}}} \sum_{k,l\in \mbB_{\vec{r}_{m}}} \ketbra{0}^{P_C} \otimes \dket{\1}^{P_T I_i}\dbra{\1}^{P_T I_j} \otimes \1^{O'_l \to O'_k} \otimes A_{ijkl}\nonumber\\
        &+ \ketbra{1}^{P_C} \otimes \dket{\1}^{P_T I'_k}\dbra{\1}^{P_T I'_l} \otimes \dket{\1}^{O'_k I_i}\dbra{\1}^{O'_l I_j} \otimes B_{ijkl},
    \end{align}
    where $A_{ijkl}$ and $B_{ijkl}$ are defined by
        \begin{align}
        A_{ijkl} &\coloneqq
        \begin{cases}
            \dket{\1}^{O_i I'_k}\dbra{\1}^{O_j I'_l} \otimes \tilde{C}_{P\vec{r}_{m}}^{(ijkl)} \otimes \1^{O_{r_{m}}} & (i,j\neq r_{m})\\
            {1\over d} C_{P\vec{r}_{m}}^{(ijkl)} \dket{\1}^{I'_k O_{r_{m}}} \dbra{\1}^{O_j I'_l} \otimes \1^{O_{r_{m}}} & (i=r_{m} \neq j)\\
            {1\over d}\dket{\1}^{I'_k O_i} \dbra{\1}^{O_{r_{m}} I'_l} C_{P\vec{r}_{m}}^{(ijkl)} \otimes \1^{O_{r_{m}}} & (j = r_{m} \neq i)\\
            {1\over d}\1^{I'_l \to I'_k} \otimes C_{P\vec{r}_{m}}^{(ijkl)} \otimes \1^{O_{r_m}}  & (i=j=r_{m})
        \end{cases},
        \label{eq:Aijkl}\\
        B_{ijkl} &\coloneqq
        \begin{cases}
            \1^{O_j \to O_i} \otimes \tilde{C}_{P\vec{r}_{m}}^{(ijkl)} \otimes \1^{O_{r_{m}}} & (i,j\neq r_{m})\\
           {1\over d}C_{P\vec{r}_{m}}^{(ijkl)} \1^{O_j \to O_{r_{m}}} \otimes \1^{O_{r_{m}}} & (i=r_{m} \neq j)\\
            {1\over d}\1^{O_{r_{m}} \to O_i} C_{P\vec{r}_{m}}^{(ijkl)} \otimes \1^{O_{r_{m}}} & (j = r_{m} \neq i)\\
            C_{P\vec{r}_{m}}^{(ijkl)} \otimes \1^{O_{r_m}} & (i=j=r_{m})
        \end{cases},
        \label{eq:Bijkl}\\
        \tilde{C}_{P\vec{r}_{m}}^{(ijkl)}&\coloneqq {1\over d}\Tr_{O_{r_{m}}}C_{P\vec{r}_{m}}^{(ijkl)}.
    \end{align}
    Using Lemma \ref{lem:linear_independence} for Eq.~\eqref{eq:apply_qccc_condition_1}, we obtain
    \begin{align}
        \sum_{k,l\in \mbB_{\vec{r}_{m}}} \dket{\1}^{O_i I'_k}\dbra{\1}^{O_j I'_l} 
        \otimes \1^{O'_l \to O'_k}
        \otimes C_{P\vec{r}_{m}}^{(ijkl)} &= \sum_{k,l\in \mbB_{\vec{r}_{m}}} 
         \1^{O'_l \to O'_k} \otimes A_{ijkl} 
        \quad \forall i,j,\label{eq:eq_for_A}\\
        \1^{O_j\to O_i} \otimes C_{P\vec{r}_{m}}^{(ijkl)} &= B_{ijkl} \quad \forall i,j,k,l.\label{eq:eq_for_B}
    \end{align}
    From Eq.~\eqref{eq:eq_for_B}, we obtain
    \begin{align}
        C_{P\vec{r}_{m}}^{(ijkl)} &= \begin{cases}
            \tilde{C}_{P\vec{r}_{m}}^{(ijkl)} \otimes \1^{O_{r_{m}}} & (i,j\neq r_{m})\\
            0 & (i = r_{m}\neq j \;\mathrm{or}\; j = r_{m}\neq i)
        \end{cases},
    \end{align}
    where the cases of $i=r_m\neq j$ and $j=r_m\neq i$ are shown as below.
    If $i=r_m\neq j$ holds, from Eqs.~\eqref{eq:Bijkl} and \eqref{eq:eq_for_B}, we obtain
    \begin{align}
        \1^{O_j\to O_{r_{m}}} \otimes C_{P\vec{r}_{m}}^{(ijkl)} = {1\over d}C_{P\vec{r}_{m}}^{(ijkl)} \1^{O_j \to O_{r_m}} \otimes \1^{O_{r_{m}}}.
        \label{i=rm_neq_j}
    \end{align}
    By taking the inner product of Eq.~\eqref{i=rm_neq_j} with $\1^{O_j \to O_{r_{m}}}$, we obtain
    \begin{align}
    \label{eq:dCPrm}
        d C_{P\vec{r}_{m}}^{(ijkl)} = {1\over d} C_{P\vec{r}_{m}}^{(ijkl)},
    \end{align}
    i.e., $C_{P\vec{r}_{m}}^{(ijkl)} = 0$ holds for $i=r_m\neq j$.
    We can similarly show that $C_{P\vec{r}_{m}}^{(ijkl)} = 0$ for $j=r_m\neq i$.
    From Eq.~\eqref{eq:eq_for_A} for $i=j=r_{m}$, we obtain
    \begin{align}
    \label{eq:i=j=rm}
        \sum_{k,l\in \mbB_{\vec{r}_{m}}} \dket{\1}^{O_{r_{m}} I'_k}\dbra{\1}^{O_{r_{m}} I'_l}
        \otimes \1^{O'_l \to O'_k} \otimes C_{P\vec{r}_{m}}^{(ijkl)} = \sum_{k,l\in \mbB_{\vec{r}_{m}}}  \1^{O'_l \to O'_k} \otimes \1^{I'_l \to I'_k} \otimes {\1^{O_{r_{m}}} \over d} \otimes C_{P\vec{r}_{m}}^{(ijkl)} \quad \mathrm{if}\; i=j=r_{m}.
    \end{align}
    Using Lemma \ref{lem:linear_independence2}, we obtain
    \begin{align}
        C_{P\vec{r}_{m}}^{(ijkl)} = 0 \quad \mathrm{if}\; i=j=r_{m}.
    \end{align}
    In conclusion, we obtain
    \begin{align}
        C_{P\vec{r}_{m}}^{(ijkl)} &= \begin{cases}
            \tilde{C}_{P\vec{r}_{m}}^{(ijkl)} \otimes \1^{O_{r_{m}}} & (i,j\neq r_{m})\\
            0 & (\mathrm{otherwise})
        \end{cases}.
    \end{align}
    Thus, from Eqs.~\eqref{eq:qc-cc_m} and \eqref{eq:induction_m+1}, we obtain
    \begin{align}
    C_{P \vec{r}_m}
        &=\sum_{i,j\in \mbA_{\vec{r}_{m-1}}} \sum_{k,l\in \mbB_{\vec{r}_{m-1}}} (\ketbra{0}^{P_C} \otimes \dket{\1}^{P_T I_i}\dbra{\1}^{P_T I_j} \otimes \dket{\1}^{O_i I'_k}\dbra{\1}^{O_j I'_l} \otimes \1^{O'_l \to O'_k} \nonumber\\
        &+ \ketbra{1}^{P_C} \otimes \dket{\1}^{P_T I'_k}\dbra{\1}^{P_T I'_l} \otimes \dket{\1}^{O'_k I_i}\dbra{\1}^{O'_l I_j} \otimes \1^{O_j \to O_i}) \otimes \tilde{C}_{P\vec{r}_{m}}^{(ijkl)}.\label{eq:C_r_i}
    \end{align}
    Thus, defining $C_{P\vec{r}_{m-1}}^{(ijkl)}$ by
    \begin{align}
        C_{P\vec{r}_{m-1}}^{(ijkl)} \coloneqq \sum_{r_{m}} \Tr_{I_{r_m}} [\tilde{C}_{P\vec{r}_{m}}^{(ijkl)}],
    \end{align}
    we obtain Eq.~\eqref{eq:C_Prm}.
\end{proof}

\vspace{3ex}

\begin{lemma}
\label{lem:linear_independence}
    The set of matrices
    \begin{align}
    \label{eq:set_A}
        \left\{\dket{\1}^{P_T I_k'} \dbra{\1}^{P_T I_l'} \otimes \dket{\1}^{O'_k I_i} \dbra{\1}^{O'_l I_j} \otimes \ket{\vec{\alpha}}^{I_{\bar{k}}'}\bra{\vec{\beta}}^{I_{\bar{l}}'} \otimes \ket{\vec{\gamma}}^{I_{\bar{i}}}\bra{\vec{\delta}}^{I_{\bar{j}}}\right\}_{\substack{i,j\in\{1, \ldots, M\}, k,l\in\{1, \ldots, N\}, \\ \vec{\alpha}, \vec{\beta}\in \{1, \ldots, d\}^{N-1}, \vec{\gamma}, \vec{\delta}\in\{1, \ldots, d\}^{M-1}}}
    \end{align}
    is linearly independent if $\max (M, N)\leq d$ holds. Similarly, the set of matrices
    \begin{align}
    \label{eq:set_B}
        \left\{\dket{\1}^{P_T I_i}\dbra{\1}^{P_T I_j} \otimes \ket{\vec{\alpha}}^{I_{\bar{i}}}\bra{\vec{\beta}}^{I_{\bar{j}}}\right\}_{\substack{i,j\in\{1, \ldots, M\}, \\ \vec{\alpha}, \vec{\beta}\in\{1, \ldots, d\}^{M-1}}}
    \end{align}
    is linearly independent if $M\leq d$ holds.
\end{lemma}

\begin{proof}
    We consider the equation
    \begin{align}
    \label{eq:linear_independence}
        \sum_{i,j,k,l,\vec{\alpha},\vec{\beta},\vec{\gamma},\vec{\delta}}A_{ijkl\vec{\alpha}\vec{\beta}\vec{\gamma}\vec{\delta}}\dket{\1}^{P_T I_k'} \dbra{\1}^{P_T I_l'} \otimes \dket{\1}^{O'_k I_i} \dbra{\1}^{O'_l I_j} \otimes \ket{\vec{\alpha}}^{I_{\bar{k}}'}\bra{\vec{\beta}}^{I_{\bar{l}}'} \otimes \ket{\vec{\gamma}}^{I_{\bar{i}}}\bra{\vec{\delta}}^{I_{\bar{j}}} = 0
    \end{align}
    for complex coefficients $A_{ijkl\vec{\alpha}\vec{\beta}\vec{\gamma}\vec{\delta}}$.
    Since $\max (M, N)\leq d$ holds, for all $\vec{\alpha}, \vec{\beta}, \vec{\gamma}, \vec{\delta}$, there exists $\alpha^*, \beta^*, \gamma^*, \delta^* \in \{1, \ldots, d\}$ such that $\alpha^*, \beta^*, \gamma^*, \delta^*$ do not appear in $\vec{\alpha}, \vec{\beta}, \vec{\gamma}, \vec{\delta}$, respectively.
    By taking an inner product of Eq.~\eqref{eq:linear_independence} with $\ket{\alpha^* \alpha^*}^{P_T I'_k}\bra{\beta^* \beta^*}^{P_T I'_l} \otimes \ket{\gamma^* \gamma^*}^{O'_k I_i} \bra{\delta^* \delta^*}^{O'_l I_j} \otimes \ket{\vec{\alpha}}^{I_{\bar{k}}'}\bra{\vec{\beta}}^{I_{\bar{l}}'} \otimes \ket{\vec{\gamma}}^{I_{\bar{i}}}\bra{\vec{\delta}}^{I_{\bar{j}}}$ for any $i,j,k,l,\vec{\alpha},\vec{\beta},\vec{\gamma},\vec{\delta}$, we obtain
    \begin{align}
   A_{ijkl\vec{\alpha}\vec{\beta}\vec{\gamma}\vec{\delta}} = 0,
    \end{align}
    i.e., the set \eqref{eq:set_A} is linearly independent.
    We can similarly show that the set \eqref{eq:set_B} is linearly independent.
\end{proof}

\vspace{3ex}

\begin{lemma}
\label{lem:linear_independence2}
    The set of matrices
    \begin{align}
    \label{eq:set_C}
        \left\{ \left(\dket{\1}^{O_{r_m} I'_k} \dbra{\1}^{O_{r_m} I'_l} - {\1^{O_{r_m}} \over d} \otimes \1^{I'_l \to I'_k} \right) \otimes \1^{O'_l \to O'_k} \otimes \ket{\vec{\alpha}}^{I'_{\bar{k}}} \bra{\vec{\beta}}^{I'_{\bar{l}}} \otimes \ket{\vec{\gamma}}^{O'_{\bar{k}}} \bra{\vec{\delta}}^{O'_{\bar{l}}} \right\}_{\substack{k,l\in\{1, \ldots, N\}, \\ \vec{\alpha}, \vec{\beta}, \vec{\gamma}, \vec{\delta}\in\{1, \ldots, d\}^{N-1}}}
    \end{align}
    is linearly independent if $N\leq \max(2, d-1)$ holds.
\end{lemma}
\begin{proof}
    We numerically check the linear independence for the case $N=d=2$ (see Listing~\ref{code}).
    We prove the linear independence for the case $N\leq d-1$ to complete the proof.
    
    We consider the equation
    \begin{align}
    \label{eq:linear_indeoendence2}
        \sum_{k,l,\vec{\alpha},\vec{\beta}, \vec{\gamma},\vec{\delta}} A_{kl\vec{\alpha}\vec{\beta}\vec{\gamma}\vec{\delta}} \left(\dket{\1}^{O_{r_m} I'_k} \dbra{\1}^{O_{r_m} I'_l} - {\1^{O_{r_m}} \over d} \otimes \1^{I'_l \to I'_k} \right) \otimes \1^{O'_l \to O'_k} \otimes \ket{\vec{\alpha}}^{I'_{\bar{k}}} \bra{\vec{\beta}}^{I'_{\bar{l}}} \otimes \ket{\vec{\gamma}}^{O'_{\bar{k}}} \bra{\vec{\delta}}^{O'_{\bar{l}}} = 0
    \end{align}
    for complex coefficients $A_{kl\vec{\alpha}\vec{\beta}\vec{\gamma}\vec{\delta}}$.
    Since $N\leq d-1$ holds, for all $\vec{\alpha}, \vec{\beta}$, there exists $\alpha^*, \beta^* \in \{1, \ldots, d\}$ such that $\alpha^*\neq \beta^*$ holds and $\alpha^*, \beta^*$ do not appear in $\vec{\alpha}, \vec{\beta}$, respectively.
    By taking an inner product of Eq.~\eqref{eq:linear_indeoendence2} with ${1\over d} \ket{\alpha^*\alpha^*}^{O_{r_m} I'_k} \bra{\beta^* \beta^*}^{O_{r_m} I'_l} \otimes \ket{\vec{\alpha}}^{I'_{\bar{k}}} \bra{\vec{\beta}}^{I'_{\bar{l}}} \otimes \1^{O'_l \to O'_k} \otimes \ket{\vec{\gamma}}^{O'_{\bar{k}}} \bra{\vec{\delta}}^{O'_{\bar{l}}}$ for any $k,l,\vec{\alpha},\vec{\beta},\vec{\gamma},\vec{\delta}$, we obtain
    \begin{align}
        A_{kl \vec{\alpha} \vec{\beta} \vec{\gamma} \vec{\delta}} = 0,
    \end{align}
    i.e., the set \eqref{eq:set_C} is linearly independent.
\end{proof}

\begin{lstlisting}[caption={MATLAB \cite{matlab} code to check the linear independency of the set \eqref{eq:set_C} for the case $d=2$ and $N=2$, which uses the functions from {\sc QETLAB} \cite{qetlab}.},label=code]
clear

d=2;
N=2;

one = Tensor(IsotropicState(d, 1)*d,eye(d));
id = Tensor(eye(d)/d,eye(d),eye(d));
I = eye(d^(d-1));

for i = 1:d
    sys(i)=i;
end
PP = perms(sys);

pos=0;

% Calculate the set of matrices
for alpha = 1:d^(d-1)
    for beta=1:d^(d-1)
        for gamma = 1:d^(d-1)
            for delta=1:d^(d-1)
                for k = 1:size(PP,1)
                    for l=1:size(PP,1)
                        pos=pos+1;
                        A(:,:,pos) = Tensor(eye(d), PermutationOperator(d, PP(k,:)), PermutationOperator(d, PP(k,:))) * PermuteSystems(Tensor(one-id, I(:,alpha)*I(beta,:), I(:,gamma)*I(delta,:)), [1 2 4 3 5]) * Tensor(eye(d), PermutationOperator(d, PP(l,:)), PermutationOperator(d, PP(l,:)));
                    end
                end
            end
        end
    end
end

% Flatten the matrices to vectors
for pos = 1:size(A,3)
    B(:,pos) = reshape(A(:,:,pos), [], 1);
end

rank(B) == size(B,2)
\end{lstlisting}

\end{document}